\documentclass[journal]{IEEEtran}

\usepackage[T1]{fontenc}
\usepackage[utf8]{inputenc}
\usepackage{amsmath,amssymb,amsthm}
\usepackage{bm}           
\usepackage{graphicx}
\usepackage{booktabs}
\usepackage{array}
\usepackage{multirow}
\usepackage{xcolor}
\usepackage{cite}
\usepackage{url}
\usepackage{tikz}
\usetikzlibrary{positioning,arrows.meta}
\usepackage[colorlinks=true,linkcolor=blue,citecolor=blue,urlcolor=blue]{hyperref}
\usepackage[font=footnotesize]{caption}
\usepackage[table]{xcolor}
\definecolor{myblue}{RGB}{217,226,243}

\usepackage{comment}

\newcommand{\bF}{\mathbf{F}}     
\newcommand{\bB}{\mathbf{B}}     
\newcommand{\bH}{\mathbf{H}}     
\newcommand{\bs}{\mathbf{s}}     
\newcommand{\Nrf}{N_{\mathrm{RF}}}   
\newcommand{\Nt}{N_t}        
\newcommand{\Nr}{N_r}        
\newcommand{\Ncl}{N_{\mathrm{cl}}}   
\newcommand{\Nray}{N_{\mathrm{ray}}} 
\newcommand{\calC}{\mathcal{C}}      

\usepackage{subcaption}
\newcommand{\etal}{\emph{et al. }}

\usepackage{tikz}
\usetikzlibrary{arrows.meta,positioning,calc}

\usepackage{balance}

\begin{document}
\setlength{\belowdisplayskip}{5pt} 
\setlength{\belowdisplayshortskip}{5pt}
\setlength{\abovedisplayskip}{5pt} 
\setlength{\abovedisplayshortskip}{5pt}

\title{Hybrid Beamforming in Non-Terrestrial Networks:\\
       Architectures, Design Challenges, and Opportunities}

\author{%
    Thuan Van Le, Nguyen Cong Luong, Huy~T.~Nguyen, \IEEEmembership{Member, IEEE}, Trong-Dai Hoang, \IEEEmembership{Graduate Student Member, IEEE}, Xiaojing Huang, \IEEEmembership{Senior Member, IEEE}, Peiyuan Qin, \IEEEmembership{Senior Member, IEEE}, 
    Tran Thien Thanh,~\IEEEmembership{Member,~IEEE},
    Vo Nguyen Quoc Bao,~\IEEEmembership{Senior Member,~IEEE}, and Ngo~Hoang~Tu,~\IEEEmembership{Member,~IEEE}
    \thanks{Thuan Van Le is with the Faculty of Electrical and Electronic Engineering, Phenikaa School of Engineering, Phenikaa University, Hanoi 12116, Vietnam (e-mail: thuan.levan@phenikaa-uni.edu.vn).}
    \thanks{Nguyen Cong Luong is with the Phenikaa School of Computing, Phenikaa University, Hanoi 12116, Vietnam (e-mail: luong.nguyencong@phenikaa-uni.edu.vn).}
    \thanks{Trong-Dai Hoang, Xiaojing Huang, and Peiyuan Qin are with the Global Big Data Technologies Centre, University of Technology Sydney, Ultimo, NSW 2007, Australia (e-mail: dai.t.hoang@student.uts.edu.au, xiaojing.huang@uts.edu.au, peiyuan.qin@uts.edu.au).}
    \thanks{Tran Thien Thanh is with the Institute of Information Technology and Electrical-Electronics Engineering, Ho Chi Minh City University of Transport, Ho Chi Minh City 700000, Vietnam (e-mail: thanh.tran@ut.edu.vn).}
    \thanks{Huy T. Nguyen, Vo Nguyen Quoc Bao and Ngo Hoang Tu are with the Faculty of Information Technology, Van Lang School of Technology, Van Lang University, Ho Chi Minh City 70000, Vietnam (e-mail: huy.nt@vlu.edu.vn, bao.vnq@vlu.edu.vn, tu.nh@vlu.edu.vn). (\textit{Corresponding author: Ngo Hoang Tu.})} 
}

\markboth{Manuscript Submitted to IEEE Communications Surveys and Tutorials, 2026}{}

\maketitle

\begin{abstract}
Hybrid analog--digital beamforming (HBF) has emerged as a key enabling technology for non-terrestrial networks (NTNs), where large antenna arrays are required to compensate for severe propagation loss but fully digital beamforming is often impractical due to radio-frequency (RF) chain cost, power consumption, and payload limitations. 
Compared with terrestrial networks, NTN platforms such as low Earth orbit (LEO) satellites and unmanned aerial vehicles (UAVs) impose distinctive HBF design challenges, including high mobility, Doppler effects, sparse line-of-sight-dominant channels, stringent on-board energy budgets, and, for UAVs, the additional coupling between beamforming and controllable platform placement or trajectory. 
This survey provides a systematic review of HBF techniques for NTN systems, with emphasis on LEO satellite and UAV communications. We first introduce common HBF architectures, signal models, channel representations, analog and digital precoder designs, and learning-aided approaches that form the shared technical foundation of existing works. 
We then survey both platforms under a common set of five categories, which cover system architecture and precoding design, time-varying beam management, network-level cooperation and scheduling, sensing capability and reconfigurable surfaces, and security and multiple access. 
Their platform-specific content differs most sharply in the second one, since the dominant time-varying mechanism is traffic-driven beam hopping on an LEO payload but mobility-aware beam tracking on a UAV.
Finally, we discuss open research challenges and future directions toward scalable, robust, and hardware-efficient HBF in next-generation NTNs.
\end{abstract}


\vspace{-0.25cm}
\section{Introduction}
\label{sec:intro}

\IEEEPARstart{N}{on-terrestrial} networks (NTNs) are becoming an essential component of future wireless systems because they can extend broadband connectivity beyond the coverage footprint of conventional terrestrial infrastructure. In particular, low Earth orbit (LEO) satellites can provide wide-area and global service coverage, while unmanned
aerial vehicles (UAVs) can be rapidly deployed as aerial base stations (BSs), relays, sensing platforms, or emergency communication nodes. Recent studies on NTN evolution and modeling have shown that aerial and spaceborne platforms introduce new opportunities for ubiquitous
coverage, resilient networking, and flexible service provisioning, but also require new radio-access and resource-management mechanisms that differ substantially from terrestrial networks~\cite{wang_non-terrestrial_2024,10969825}.

Beamforming is one of the most important physical-layer technologies for
NTNs. Satellite and UAV links often require highly directional transmission
and large-scale antenna arrays to compensate for severe propagation loss over
long distances. Meanwhile, millimeter-wave (mmWave) and terahertz (THz)
bands are attractive because they provide large transmission bandwidths and
allow electrically large arrays to be integrated within compact physical
apertures, although they also introduce more challenging propagation and
hardware conditions. However, fully digital beamforming requires one
dedicated radio-frequency (RF) chain per antenna element.
This architecture provides high spatial flexibility but becomes costly, power-hungry, and difficult to implement on size-, weight-, and power-constrained NTN platforms. Hybrid analog--digital beamforming (HBF) addresses this limitation by combining a low-dimensional digital precoder with an analog RF beamforming network, thereby reducing the number of RF chains while retaining a large portion of the array gain and spatial multiplexing capability of fully digital systems~\cite{ahmed_survey_2018}.

Although HBF has been extensively studied for terrestrial massive multiple-input multiple-output (MIMO) and mmWave systems, its use in NTNs is not a straightforward extension. LEO satellites move at orbital speeds, which creates large Doppler shifts, rapidly changing beam footprints, and non-negligible channel aging. Consequently, LEO HBF must often be designed together with beam management, beam hopping, user scheduling, statistical or predicted channel state information (CSI), and payload-aware hardware constraints~\cite{you_hybrid_2022,you_beam_2022,liu_joint_2022,Zhang2022LEOChannelPrediction}.
UAV systems exhibit a different set of couplings: the aerial platform location, altitude, trajectory, mechanical jitter, energy budget, and relay topology can directly affect the beamforming gain and interference pattern~\cite{du_energy-saving_2020,zhu_multi-uav_2022,chen_adaptive_2024,liu_deployment_2023}.
These differences make NTN HBF a distinct research topic requiring a dedicated survey.

The resulting NTN HBF literature spans several tightly coupled research
directions. Satellite-oriented works have studied statistical-CSI-based HBF, beam-squint-aware HBF, beam hopping, multi-satellite cooperation, robust designs under CSI uncertainty, and fairness-aware HBF under limited feedback
\cite{palacios_hybrid_2021,you_hybrid_2022,you_beam_2022,liu_robust_2022,meng_joint_2025,11575120,11480864}.
In parallel, UAV-oriented studies have investigated energy-aware HBF,
joint trajectory and beamforming optimization, multi-UAV cooperation,
relay-assisted transmission, sensing-assisted beamforming, and secure or
multiple-access-enabled designs
\cite{feng_hybrid_2021,chen_joint_2022,wang_cooperative_2023,zhang_sensing-assisted_2024,dong_joint_2025}.
However, the value of a particular HBF design cannot be assessed solely
from its optimization objective or algorithmic label. It also depends on
how the selected transceiver architecture interacts with platform-dependent
mobility, CSI quality, RF-chain availability, payload constraints, and
resource-allocation timescales. This observation motivates a platform-aware
and architecture-aware survey of HBF for NTN systems.

\subsection{Related Surveys and Research Gap}
\label{ssec:related_surveys_gap}

A substantial body of survey literature has established the foundations of
beamforming and HBF for terrestrial wireless systems. Early
reviews on mmWave beamforming and massive-MIMO HBF have discussed analog,
digital, and hybrid transceiver architectures, sparse channel exploitation,
codebook design, and the associated complexity--performance tradeoffs
\cite{kutty2015beamforming,molisch2017hybrid}.
Among these foundational works, \cite{ahmed_survey_2018} provides a comprehensive treatment of HBF architectures and algorithms for terrestrial 5G massive MIMO and mmWave systems, and remains a standard reference for HBF signal models, analog/digital precoder structures, and architecture-level tradeoffs, whereas \cite{rihan2020taxonomy} offers a taxonomy-oriented and performance-based perspective on HBF for 5G-and-beyond systems.
More recent reviews have expanded the discussion toward vertical or
application-specific settings, including mmWave HBF for wireless intelligent
transport systems, massive-MIMO HBF design advances, and smart beamforming
for high-speed and energy-efficient wireless systems
\cite{shahjehan2024review,srivastava2025advancements,tala2026smart}.

Collectively, these surveys provide important architectural and algorithmic
foundations. Nevertheless, their primary focus remains terrestrial or
application-specific. They do not systematically address the joint impact of orbital mobility, large Doppler shifts, time-varying satellite footprints, beam hopping, payload-limited RF hardware, UAV trajectory coupling, platform jitter, and rapidly aging CSI. In particular, the existing HBF surveys do not provide a common framework for comparing how identical architectural choices, such as fully-connected, partially-connected, or dynamic-subarray HBF, should be adapted to the markedly different operating regimes of LEO satellites and UAV platforms.

Complementary survey efforts have also examined satellite communications and general NTN design. The survey in \cite{khammassi_precoding_2023} reviews precoding for high-throughput satellite systems, with emphasis on multibeam interference mitigation, robust precoding, and predominantly digital techniques. That work is highly relevant to the satellite side of NTN beamforming, but it does not focus on hybrid analog--digital architectures and does not cover UAV HBF.
Meanwhile, \cite{wang_non-terrestrial_2024} provides a broad perspective on
NTN evolution, radio resource management, mobility management, and network slicing, whereas \cite{10969825} focuses on NTN modeling and analysis through spherical stochastic geometry.
These works are valuable for understanding satellite-system operation,
network-level NTN integration, and analytical modeling. However, none of them provides a systematic review of HBF architectures, algorithms, or hardware-aware transceiver design that jointly covers LEO satellite and UAV communication systems.

The previous discussion reveals a clear gap in the current survey literature.
To the best of our knowledge, no existing survey jointly reviews HBF for LEO satellite and UAV communication systems under a common platform-aware, architecture-aware, and methodology-aware framework. These platforms represent complementary NTN regimes: high-mobility spaceborne systems with deterministic orbital motion and payload constraints, and controllable aerial systems affected by trajectory, jitter, and battery limitations. Medium Earth orbit (MEO) and geostationary Earth orbit (GEO) satellites and high altitude platform stations (HAPSs) are discussed where their HBF principles extend the main LEO--UAV framework. A unified treatment is warranted because LEO and UAV systems share sparse directional channels, limited RF-chain architectures,
analog beam codebooks, and reduced-dimensional digital precoding, yet differ markedly in mobility, CSI acquisition, hardware constraints, and cross-layer coupling.

\begin{table*}[t]
\centering
\caption{\normalfont\scshape Comparison of Related Surveys and Position of This Survey}
\label{tab:existing_surveys}
\renewcommand{\arraystretch}{1.18}
\setlength{\tabcolsep}{2.5pt}
\scriptsize

\begin{tabular}{|>{\raggedright\arraybackslash}p{1.5cm}|
p{3.00cm}|
p{2.55cm}|
p{2.80cm}|
p{7.2cm}|}
\hline
\cellcolor{myblue}\textbf{Survey} &
\cellcolor{myblue}\textbf{Main Focus} &
\cellcolor{myblue}\textbf{Primary Platforms} &
\cellcolor{myblue}\textbf{Beamforming Scope} &
\cellcolor{myblue}\textbf{Gap Relative to This Survey} \\
\hline
\hline

\multicolumn{5}{|l|}{\cellcolor{gray!15}\textit{Foundational HBF and terrestrial/vertical-domain surveys}} \\
\hline

Kutty and Sen \cite{kutty2015beamforming} &
Beamforming techniques for mmWave communications &
Terrestrial mmWave systems &
Analog, digital, and HBF &
Provides broad mmWave beamforming foundations, but does not consider NTN-specific mobility, satellite payload constraints, beam hopping, or UAV trajectory coupling. \\
\hline

Molisch \emph{et al.} \cite{molisch2017hybrid} &
HBF for massive MIMO systems &
Terrestrial massive MIMO and mmWave systems &
HBF architectures, channel estimation, and precoding &
Establishes key HBF principles, but does not provide platform-aware design guidance for LEO satellites and UAV systems. \\
\hline

Ahmed \emph{et al.} \cite{ahmed_survey_2018} &
HBF for 5G massive MIMO and mmWave systems &
Terrestrial cellular networks &
Hybrid analog--digital transceivers and algorithms &
Provides a foundational terrestrial HBF survey; LEO Doppler, beam hopping, payload limitations, UAV jitter, and trajectory--beamforming coupling are outside its scope. \\
\hline

Rihan \emph{et al.} \cite{rihan2020taxonomy} &
Taxonomy and performance evaluation of HBF for 5G and beyond &
Terrestrial 5G/B5G systems &
HBF taxonomy and performance evaluation &
Presents a useful HBF taxonomy, but it is not tailored to the distinct mobility, CSI, and hardware constraints of LEO and UAV platforms. \\
\hline

Shahjehan \emph{et al.} \cite{shahjehan2024review} &
mmWave HBF for wireless intelligent transport systems &
Terrestrial vehicular and transport systems &
mmWave HBF for intelligent transport systems &
Targets road and vehicular communication environments rather than aerial or spaceborne platforms, and does not address LEO/UAV HBF. \\
\hline

Srivastava \emph{et al.} \cite{srivastava2025advancements} &
Recent advances, challenges, and opportunities in massive-MIMO HBF &
Terrestrial massive MIMO systems &
HBF design trends and implementation challenges &
Reviews massive-MIMO HBF developments, but does not study NTN-specific channel aging, platform mobility, payload constraints, or beam-management coupling. \\
\hline

Tala \emph{et al.} \cite{tala2026smart} &
Smart beamforming for high-speed and energy-efficient wireless systems &
General high-speed wireless systems &
Smart beamforming techniques &
Provides a broad smart-beamforming perspective, but is not a dedicated survey of HBF for LEO satellites and UAV systems. \\
\hline
\hline

\multicolumn{5}{|l|}{\cellcolor{gray!15}\textit{Satellite and NTN-oriented surveys}} \\
\hline

Khammassi \emph{et al.}~\cite{khammassi_precoding_2023} &
Precoding for high-throughput satellite communication systems &
Multibeam satellite and high-throughput satellite (HTS) systems &
Mainly digital and satellite precoding &
Focuses on satellite precoding and multibeam interference mitigation. Hybrid analog--digital architectures and UAV HBF are not jointly addressed. \\
\hline

Wang \emph{et al.} \cite{wang_non-terrestrial_2024} &
NTN evolution, opportunities, management, and future directions &
Multi-layer NTN systems with satellite emphasis &
Network-level NTN design and management &
Provides a broad NTN perspective, but does not systematically compare HBF architectures, transceiver algorithms, or hardware-aware HBF design. \\
\hline

Wang \emph{et al.} \cite{10969825} &
NTN modeling and analysis using spherical stochastic geometry &
Satellite and aerial NTN systems &
Spatial modeling and coverage analysis &
Develops an NTN analytical-modeling perspective rather than a transceiver-level review of HBF architectures, optimization, robustness, and learning. \\
\hline
\hline

\rowcolor{gray!20}
\textbf{This survey} &
HBF in non-terrestrial networks &
LEO satellite and UAV systems, with selected aerial--space extensions &
HBF architectures, algorithms, and cross-cutting extensions &
Provides a unified platform-aware taxonomy for LEO and UAV HBF, compares architectures and design methods, and connects beam hopping, CSI prediction, robust optimization, learning, ISAC, RIS/RHS/ITSs, THz, security, and multiple access. \\
\hline

\end{tabular}
\vspace{-0.25cm}
\end{table*}

This article fills this gap by jointly surveying LEO and UAV HBF.
Accordingly, this survey treats LEO satellite and UAV systems as the two
primary platform axes and organizes the literature according to HBF
architecture, design methodology, and functional extension. Rather than
presenting robust optimization, learning-based design, integrated sensing and communications (ISAC), reconfigurable intelligent surfaces (RISs),
reconfigurable holographic surfaces (RHSs), intelligent transmitting surfaces (ITSs), THz communications, security, and multiple access as isolated topics, we position them as cross-cutting directions that modify or extend the underlying HBF design. Table~\ref{tab:existing_surveys} summarizes the related surveys discussed above and further highlights the distinction of this work.

\subsection{Survey Scope and Taxonomy}
\label{ssec:scope_taxonomy}

This survey adopts a focused rather than exhaustive platform-by-platform
scope. LEO satellites and UAVs are selected as the two primary axes because
they represent complementary spaceborne and aerial HBF regimes and provide
sufficiently distinct bodies of literature for systematic comparison.
{MEO and GEO satellites and HAPSs are not treated as separate platform categories, for three reasons that concern HBF specifically rather than satellite communications in general. First, the hardware pressure that motivates HBF is much weaker on these platforms. HBF exists because a large aperture must be driven by far fewer RF chains than antenna elements under a hard payload budget. A GEO payload operates with a kilowatt-class direct current (DC) power budget and is not mass-produced, and its multibeam coverage is conventionally synthesized by a feed cluster illuminating a large reflector, with the beamforming network optionally relocated to the ground over the feeder link. The analog--digital split is therefore an implementation option rather than a binding constraint, whereas an LEO payload and a UAV payload must both realize a large array under a strict size, weight, and power (SWaP) or battery limit. Second, the mobility that shapes NTN HBF is largely absent. A GEO satellite is stationary with respect to the served region, so its footprint, Doppler shift, and angular geometry are essentially time-invariant and beam management reduces to traffic-driven allocation over fixed beams, while an MEO visibility window lasts hours rather than the few minutes of an LEO pass. Channel aging, Doppler-aware precoding, and beam-hopping timescales, which are precisely the effects that make LEO HBF distinct, thus lose their significance. Consistently, the corresponding literature is dominated by digital multibeam precoding and interference mitigation, which is already surveyed in~\cite{khammassi_precoding_2023}, rather than by hybrid transceiver design. Third, a HAPS occupies an intermediate regime whose HBF problems are inherited rather than new, because a quasi-stationary platform at an altitude of approximately 20~km combines the station-keeping drift and the platform jitter of an aerial node with a satellite-like footprint, and is therefore covered by the mechanisms reviewed in Sections~\ref{sec:leo} and~\ref{sec:uav}. These platforms consequently enter the survey where they genuinely change the HBF problem, namely in multi-layer coordination, inter-layer backhaul, and beam or platform handover, as discussed in Section~\ref{ssec:3dntn}.}
RISs, ISAC, and THz communications are instead treated as cross-cutting enabling technologies applicable to multiple NTN platforms.

\begin{figure*}[t]
\centering
\includegraphics[width=\linewidth]{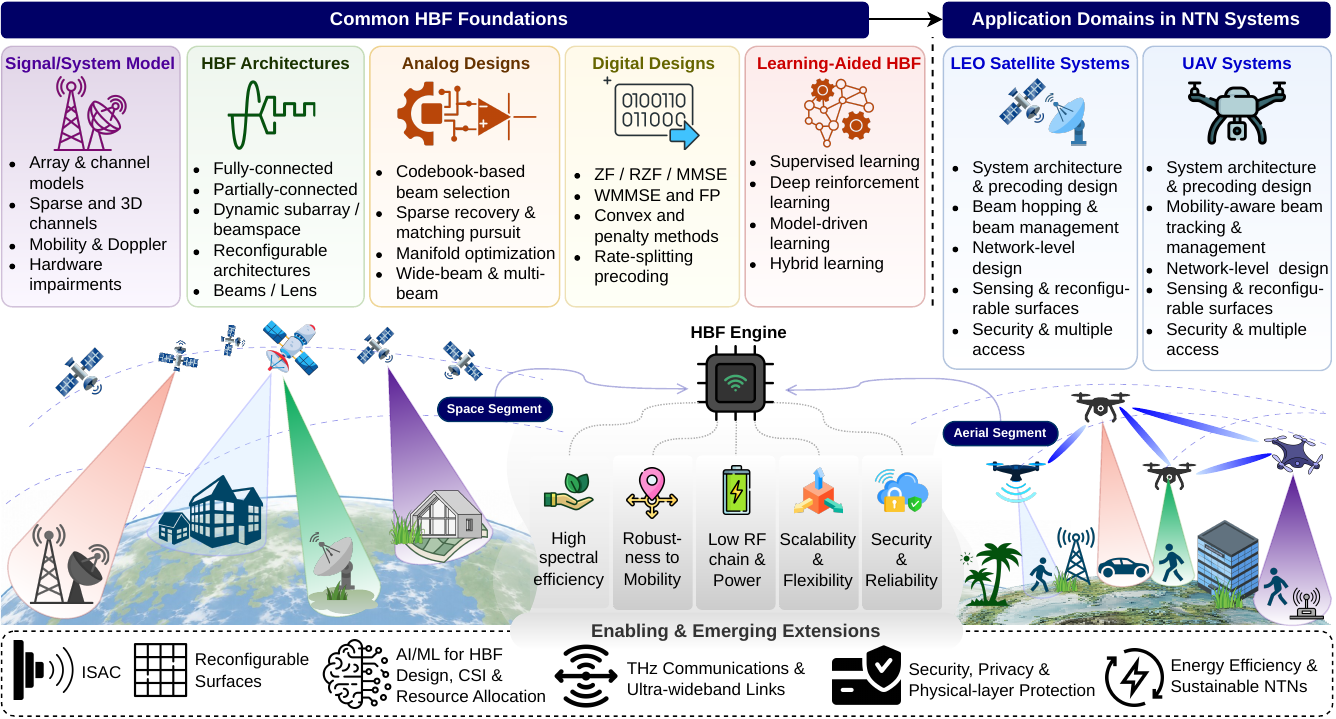}
\caption{Taxonomy of HBF in NTNs adopted in this survey.}
\label{fig:taxonomy_hbf_ntn}
\vspace{-0.25cm}
\end{figure*}

Fig.~\ref{fig:taxonomy_hbf_ntn} summarizes the taxonomy adopted in this article. Instead of organizing the literature only by algorithmic labels, we structure it in two stages. The first stage gathers the common HBF foundations shared by all non-terrestrial platforms, namely the signal and system model, the hybrid architecture families, the analog beamformer designs, the digital precoder designs, and the learning-aided methods. The second stage maps these foundations onto the two application domains that are the focus of this survey, namely LEO satellite and UAV systems, within which the same foundations are specialized and extended. This organization is chosen because the novelty of an NTN HBF paper often arises from the coupling between a shared foundational tool and a platform-specific constraint. For instance, beam hopping is mainly a satellite-specific spatio-temporal resource-allocation problem, whereas joint trajectory--beamforming design is mainly a UAV-specific mobility-control problem. 
The two stages are linked by a common HBF engine that maps the shared foundations to the design objectives, namely high spectral efficiency, mobility robustness, low RF-chain and power cost, scalability, and security. Robust and learning-based methods, together with the enabling extensions in the figure, i.e., ISAC, reconfigurable surfaces, AI/ML-aided design, THz links, physical-layer security, and energy efficiency, cut across both domains and are treated as cross-cutting rather than primary structure.

\subsection{Main Contributions}
\label{ssec:contributions}

Building on the research gap identified in Section~\ref{ssec:related_surveys_gap} and the taxonomy of Section~\ref{ssec:scope_taxonomy}, the main contributions of this survey are summarized as follows:
\begin{itemize}
    \item We provide a systematic review of HBF techniques for NTN communications, with particular emphasis on LEO satellite and UAV systems. This focus distinguishes our survey paper from generic HBF surveys, generic NTN surveys, and satellite-only
    precoding surveys.

    \item We propose a taxonomy that jointly accounts for NTN platform type, HBF architecture, design methodology, and functional extension. This taxonomy clarifies which research problems are shared by LEO and UAV systems and which are platform-specific.

    \item We review the common signal models, sparse channel representations, analog beamformer design tools, digital precoder optimization frameworks, and learning-based approaches that form the methodological foundation of NTN HBF.


    \item {We organize the LEO and UAV literature under a single set of five categories, namely (\textit{i}) system architecture and precoding design, (\textit{ii}) time-varying beam management, (\textit{iii}) network-level design, (\textit{iv}) sensing capability and reconfigurable surfaces, and (\textit{v}) security and multiple access. Within this common structure, category (\textit{ii}) covers beam hopping on LEO payloads and mobility-aware beam tracking on UAVs, category (\textit{iii}) covers multi-satellite cooperation, user scheduling, and CSI acquisition on LEO and multi-UAV cooperation and relaying on UAVs, and category (\textit{v}) additionally covers interference mitigation and spectrum coexistence on LEO. This alignment allows the two platforms to be compared category by category rather than only within a platform.}

    \item We identify key open problems in real-time implementation, CSI acquisition, Doppler- and jitter-aware design, multi-layer NTN coordination, beam-hopping integration, ISAC, THz, and learning-aided HBF.
\end{itemize}

\subsection{Organization of the Survey}
\label{ssec:organization}

The remainder of this article is organized as follows.
Section~\ref{sec:tutorial} introduces common HBF techniques used in both
satellite and UAV systems, including signal models, channel representations, analog and digital precoding, and machine-learning-aided
HBF. 
Section~\ref{sec:leo} surveys HBF for LEO satellite systems under five categories, namely (i) system architecture and precoding design, (ii) beam hopping and spatio-temporal beam management, (iii) network-level design, (iv) sensing capability and reconfigurable surfaces, and (v) security, multiple access, and interference mitigation. Section~\ref{sec:uav} reviews HBF for UAV systems under the same five categories, with (ii) addressing mobility-aware beam tracking and (iii) addressing multi-UAV cooperation and relaying.
Section~\ref{Sect:OpenChallenges} then discusses open research challenges and future directions, spanning hardware-aware HBF, wideband beam-squint control, real-time AI-assisted beam tracking, ISAC-HBF, 3D multi-layer NTN coordination, and THz HBF. 
Finally, Section~\ref{Sect:Conclusion} concludes the survey.


\section{Fundamentals of Common HBF Techniques in Satellite and UAV Networks}
\label{sec:tutorial}
This section reviews the fundamental HBF concepts that are common to satellite and UAV studies.  
The objective is not to repeat the extensive terrestrial HBF literature, but to establish a shared language for the subsequent LEO and UAV sections.  
In both platforms, the same basic transceiver decomposition is used: a low-dimensional digital precoder creates spatial streams in baseband, while an analog RF network steers high-gain beams through a large antenna array.
The specific difficulty in NTN systems is that this decomposition must be designed under mobility, sparse line-of-sight (LoS)-dominant channels, payload or battery limitations, and imperfect CSI.  
%
Table~\ref{tab:hbf_architecture_techniques} previews the architectures and design tools treated in this section and records, for each one, what it offers on an LEO payload and on a UAV. The remainder of the section develops these entries in turn.

\begin{table*}[t]
\centering
\caption{\normalfont\scshape Summary of Representative HBF Architectures and Design Techniques for NTN Systems}
\label{tab:hbf_architecture_techniques}
\renewcommand{\arraystretch}{1.22}
\setlength{\tabcolsep}{3pt}
\scriptsize

\begin{tabular}{|p{1.8cm}|p{2.8cm}|p{4.8cm}|p{3.75cm}|p{3.75cm}|}
\hline
\cellcolor{myblue} {\textbf{Architecture / Technique}} &
\cellcolor{myblue} {\textbf{Main Principle}} &
\cellcolor{myblue} {\textbf{Typical Design Tools}} &
\cellcolor{myblue} {\textbf{Relevance to LEO Satellites}} &
\cellcolor{myblue} {\textbf{Relevance to UAV Systems}} \\
\hline

 {\textbf{Fully-connected HBF}} &
 {Each RF chain is connected to all antennas through phase shifters.} &
 {Alternating optimization (AO), manifold optimization, weighted minimum mean-square error (WMMSE), orthogonal matching pursuit (OMP), and successive convex approximation (SCA)~\cite{ahmed_survey_2018,you_hybrid_2022}.} &
 {High array gain and flexible multibeam precoding, but high payload power and calibration cost.} &
 {High spectral efficiency for UAV-BS links, but heavy RF power consumption.} \\
\hline

 {\textbf{Partially-connected HBF}} &
 {Each RF chain feeds a subarray, yielding a block-diagonal analog precoder.} &
 {Subarray assignment, minimum mean-square error (MMSE)/zero-forcing (ZF)} baseband precoding, and Riemannian updates~\cite{you_massive_2022,chen_joint_2022}. &
 {Attractive for satellite terminals and low-complexity payloads, but with lower interference-management capability.} &
 {Suitable for compact UAVs with limited battery and payload capacity.} \\
\hline

 {\textbf{Dynamic subarray / beamspace HBF}} &
 {RF chains are connected to selected beams or adaptive subarrays.} &
 {Lens-array beam selection, penalty dual decomposition (PDD)}, sparse recovery, and codebook search~\cite{palacios_dynamic_2022,chen_hybrid_2021,chen_joint_2022}. &
 {Supports moving footprints and reduces RF-chain usage in sparse channels.} &
 {Useful for UAV mmWave links with angular sparsity and altitude-dependent coverage.} \\
\hline

 {\textbf{Holographic / RIS / ITS-assisted HBF}} &
 {A reconfigurable surface shapes the wavefront before or after the RF network.} &
 {Surface coefficient optimization, gradient methods, semidefinite relaxation (SDR)/SCA, and statistical-CSI designs~\cite{deng_holographic_2022,li_holographic_2025,sheemar_joint_2026,wu_hybrid_2023}.} &
 {Promising for low-profile user terminals and beam-squint mitigation in wideband NTN links.} &
 {Enables coverage extension and secure indoor--outdoor aerial links.} \\
\hline

 {\textbf{Analog beam selection}} &
 {Selects beams from a predefined or dynamic codebook.} &
 {Discrete Fourier transform (DFT)}/hierarchical codebooks, location-aided selection, and dynamic codebook updates~\cite{palacios_hybrid_2021,palacios_dynamic_2022,momani_beam_2025}. &
 {Well matched to predictable satellite geometry and slow analog-beam update rates.} &
 {Useful when location or sensing information can replace full CSI feedback.} \\
\hline

 {\textbf{Digital effective-channel precoding}} &
 {Designs the baseband precoder over the reduced channel $\tilde{\bH}=\bH\bF$.} &
 {ZF, regularized ZF (RZF), MMSE, WMMSE, and fractional programming (FP)}~\cite{arnau_performance_2012,zhai_hybrid_2021,miao_serving_2025}. &
 {Handles multibeam interference and multicast/user scheduling with reduced dimension.} &
 {Supports multi-user interference suppression after analog beam alignment.} \\
\hline

 {\textbf{Learning-aided HBF}} &
 {Learns CSI prediction, beam selection, or direct precoder generation.} &
 {Deep neural network (DNN)/convolutional neural network (CNN), deep reinforcement learning (DRL), multi-agent proximal policy optimization (MAPPO)}, and model-driven unfolding~\cite{Zhang2022LEOChannelPrediction,ren_machine_2019,meng_joint_2025,hevesli_hybrid_2025}. &
 {Promising for CSI aging, beam hopping, and predictable orbital dynamics.} &
 {Promising for beam tracking, trajectory coupling, and fast adaptation to jitter.} \\
\hline

\end{tabular}
\vspace{-0.25cm}
\end{table*}

\subsection{System Model and HBF Architectures}
\label{ssec:sysmodel}

\subsubsection{Signal model}
Consider a downlink system in which a non-terrestrial transmitter equipped with
$\Nt$ antennas and $\Nrf$ RF chains serves $K$ single-antenna users.  The condition
$\Nrf \ll \Nt$ captures the defining motivation of HBF: the antenna array is large,
but only a small number of expensive RF chains is available.  The transmitted signal
is given by
\begin{equation}
  \mathbf{x} = \bF \bB \bs,
  \label{eq:tx}
\end{equation}
where $\bF \in \mathbb{C}^{\Nt \times \Nrf}$ is the analog precoder implemented in the
RF domain, $\bB \in \mathbb{C}^{\Nrf \times K}$ is the digital baseband precoder, and
$\bs \in \mathbb{C}^{K \times 1}$ is the data-symbol vector with
$\mathbb{E}\{\bs\bs^H\}=\mathbf{I}$.  The received signal at user $k$ is expressed as
\begin{equation}
  y_k = \mathbf{h}_k^H \bF \bB \bs + n_k,
  \label{eq:rx_user}
\end{equation}
where $\mathbf{h}_k \in \mathbb{C}^{\Nt \times 1}$ is the downlink channel and
$n_k \sim \mathcal{CN}(0,\sigma^2)$ is additive white Gaussian noise.  If the
$k$-th data stream is precoded by the $k$-th column $\mathbf{b}_k$ of $\bB$, the achievable
rate is commonly expressed as
\begin{equation}
  R_k = \log_2\!\left(1+
    \frac{|\mathbf{h}_k^H \bF \mathbf{b}_k|^2}
         {\sum_{j\neq k}|\mathbf{h}_k^H \bF \mathbf{b}_j|^2 + \sigma^2}
  \right),
  \label{eq:rate_user}
\end{equation}
with the total transmit-power constraint
\begin{equation}
    \|\bF\bB\|_F^2 \leq P_{\max}.
    \label{eq:power_constraint}
\end{equation}
For a multi-antenna receiver, the same structure is extended by adding an analog
combiner and a digital combiner.  Most LEO downlink and UAV base-station works, however, can be understood from the transmit-side model in
\eqref{eq:tx}--\eqref{eq:power_constraint}.

\begin{figure}[t]
\centering
\includegraphics[width=\linewidth]{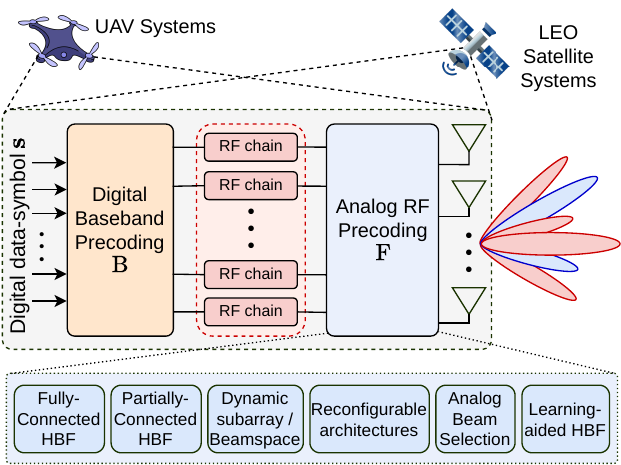}
\caption{Generic hybrid analog--digital beamforming architecture for NTN transmitters.  The digital precoder operates over a small number of RF chains, whereas the analog RF network maps these RF-chain outputs to a large antenna array.  The same abstraction is used in LEO satellite and UAV systems, but the dominant impairments differ: orbital Doppler, beam hopping, and payload constraints for LEO satellites, against trajectory coupling, battery limitations, and platform jitter for UAVs.}
\label{fig:generic_hbf_architecture}
\vspace{-0.25cm}
\end{figure}

Fig.~\ref{fig:generic_hbf_architecture} maps this model onto the hardware
implementation: $\bB$ performs baseband precoding across the $\Nrf$ RF
chains, while $\bF$ maps the RF-chain outputs to the $\Nt$ antenna elements.
The transmitted signal therefore depends on the combined precoder
$\bF\bB$. 
Two consequences follow, and they organize the rest of this section. 
First, the constraint set of $\bF$ is fixed by the RF wiring rather than by the optimization problem, which is why the architecture must be chosen before the algorithm. 
Second, the same abstraction conceals very different impairments on the two platforms: orbital Doppler, beam hopping, and payload power limits for LEO satellites, against trajectory coupling, battery endurance, and platform jitter for UAVs. 
This is why one algebraic model supports the two distinct literatures reviewed in Sections~\ref{sec:leo} and~\ref{sec:uav}.

\subsubsection{Hardware architectures}
The HBF architecture determines both the feasible set of $\bF$ and the hardware cost. The fully-connected (FC) and partially-connected (PC) structures are the two dominant architectures in the HBF literature~\cite{ahmed_survey_2018}.  
In an FC architecture, each RF chain is connected to every antenna element through a phase shifter, which gives $\bF$ a dense constant-modulus structure, typically $|[\bF]_{i,j}|=1/\sqrt{\Nt}$.  This architecture maximizes analog beamforming freedom and is attractive when spectral efficiency is the main objective.  Its disadvantage is the large number $\Nt\Nrf$ of phase shifters, which increases circuit power, calibration complexity, insertion loss, and thermal burden.

In a PC architecture, each RF chain drives only a disjoint subarray, leading to a block-diagonal analog precoder.  The number of phase shifters is reduced to $\Nt$, and hence the architecture is more suitable for payload- or battery-constrained platforms. The price is a reduced ability to form multiple overlapping beams and to manage inter-user interference.  This tradeoff appears repeatedly in both LEO and UAV studies: FC structures are preferred for high-capacity multibeam operation, whereas PC structures are attractive for low-power terminals, compact payloads, and holographic or lens-based implementations~\cite{you_hybrid_2022,you_massive_2022,chen_joint_2022,liu_deployment_2023}.

A third family includes dynamic subarrays, lens/beamspace architectures, holographic or reconfigurable surfaces, and ITSs.  These architectures attempt to retain some flexibility of FC HBF while approaching the lower hardware cost of PC HBF.  For example, beamspace HBF exploits angular sparsity through a lens array in UAV mmWave systems~\cite{chen_joint_2022,chen_hybrid_2021}, whereas holographic and reconfigurable surfaces have recently been explored for LEO satellite and UAV links~\cite{deng_holographic_2022,li_holographic_2025,sheemar_joint_2026,wu_hybrid_2023}.

Table~\ref{tab:hbf_architecture_techniques} summarizes the main HBF
architectures together with their representative design tools. No single
architecture is preferable under all operating conditions. The FC
architecture provides high array gain and flexible multibeam transmission,
but requires many phase shifters and consumes more RF power. The PC
architecture reduces hardware complexity and power consumption, but its
block-diagonal analog precoder offers less flexibility for interference
suppression. Beamspace and holographic architectures can further reduce the
required RF-chain or phase-shifter resources by exploiting angular sparsity
or a reconfigurable aperture, although their effectiveness depends strongly
on the channel structure and hardware configuration. The platform-specific
comparisons in the last two columns also show that the same architecture is
evaluated under different criteria. Satellite designs emphasize payload
power, calibration effort, and long-term hardware reliability, whereas UAV
designs place greater importance on battery endurance, payload weight, and
robustness to trajectory variations and platform jitter. Consequently, an
HBF architecture developed for one platform must be reassessed before being
applied to the other.

\subsubsection{Design implications for NTN}
The architecture choice should be made jointly with the platform constraints rather than treated as a purely signal-processing decision.  A satellite payload is launched with fixed hardware and is difficult to repair or recalibrate, so robustness to phase quantization, nonlinear power amplifiers, and thermal variations is crucial~\cite{you_massive_2022,mahmood_novel_2024}.  
A UAV payload, in contrast, is physically accessible but is constrained by flight time, vibration, and the coupling between positioning and communication performance~\cite{du_energy-saving_2020,liu_deployment_2023}.  Hence, the same HBF architecture can be preferred for different reasons: PC HBF can reduce satellite payload power, while in UAV systems it can extend battery life and simplify mechanical integration.

\subsection{Channel Models for NTN}
\label{ssec:channel}

\subsubsection{Geometric sparse channel model}
At mmWave and sub-THz frequencies, both satellite and UAV channels are often dominated
by a small number of paths.  The Saleh--Valenzuela geometric model is therefore widely
used as a common abstraction~\cite{ahmed_survey_2018}, i.e.,
\begin{equation}
  \bH = \sqrt{\frac{\Nt \Nr}{\Ncl \Nray}}
        \sum_{c=1}^{\Ncl}\sum_{r=1}^{\Nray}
        \alpha_{c,r}\,
        \mathbf{a}_r(\phi_{c,r}^{r})\,
        \mathbf{a}_t^H(\phi_{c,r}^{t}),
  \label{eq:channel}
\end{equation}
where $\alpha_{c,r}$ is the complex gain of the $r$-th ray in cluster $c$,
$\phi_{c,r}^{t}$ and $\phi_{c,r}^{r}$ are the angles of departure (AoD) and arrival (AoA), and $\mathbf{a}_t(\cdot)$ and $\mathbf{a}_r(\cdot)$ are the transmit and receive array response vectors (ARVs).  
For a uniform linear array (ULA) with half-wavelength spacing, the ARV is expressed as
\begin{equation}
  \mathbf{a}(\phi)=\frac{1}{\sqrt{\Nt}}
  \left[1,e^{j\pi\sin\phi},\ldots,e^{j\pi(\Nt-1)\sin\phi}\right]^T.
  \label{eq:ula_response}
\end{equation}
For simplicity, \eqref{eq:ula_response} illustrates only the transmit ARV and the receive ARV is defined analogously by replacing $N_t$ with the number of receive antennas.
Uniform planar arrays (UPAs) extend this representation to azimuth and elevation, which is important because both LEO satellites and UAVs require three-dimensional beam steering.

\subsubsection{NTN-specific channel features}
Three channel properties distinguish NTN HBF from conventional terrestrial HBF.  First, LoS dominance produces sparse or near-rank-one channels. This property is beneficial for analog beamforming because the dominant channel subspace can often be approximated by a small number of steering vectors.  It also motivates compressed sensing and beamspace channel estimation in LEO systems~\cite{palacios_hybrid_2021,ku_leveraging_2026}.

Second, wideband operation causes beam squint.  Since analog phase shifters create a frequency-independent phase shift, the beam direction varies across orthogonal frequency-division multiplexing (OFDM) subcarriers.
Beam squint is especially problematic for large apertures and wideband LEO downlinks, where the same analog beam must serve users over a broad bandwidth~\cite{you_beam_2022,wu_hybrid_2023}.  
UAV systems also experience beam squint at wide mmWave or THz bandwidths, but the dominant impairment is often beam misalignment caused by mobility or jitter.

Third, the time variation has different origins in LEO and UAV links.  LEO satellites move according to deterministic orbital mechanics, which creates high Doppler and predictable footprint evolution.  
UAV links are lower-speed but are affected by trajectory changes, attitude variations, and mechanical jitter. 
Therefore, LEO HBF favors statistical or predicted CSI and scheduled beam updates, whereas UAV HBF often exploits location information, sensing, or trajectory control~\cite{Zhang2022LEOChannelPrediction,chen_adaptive_2024,liu_deployment_2023,zhang_sensing-assisted_2024}.

Notably, the geometric sparsity in \eqref{eq:channel} is the structural reason HBF is effective in NTN systems.  The analog stage approximates the dominant angular subspace, while the digital stage handles residual multi-user interference over a reduced-dimensional effective channel.  The design difficulty is not the existence of sparsity, but how to exploit it when the angular support is outdated, uncertain, or coupled with platform motion.

\subsection{Analog Beamformer Design}
\label{ssec:analog}

The analog precoder $\bF$ is constrained by the RF hardware.  With phase-shifter
networks, a common constraint is
\begin{equation}
    |[\bF]_{i,j}|={1}/{\sqrt{\Nt}}, \ \forall i,j,
    \label{eq:constant_modulus}
\end{equation}
which makes the design non-convex.  In switch- or lens-based networks, additional
sparsity or selection constraints appear.  Four design families are most common.

\subsubsection{Codebook-based beam selection}
A codebook $\calC=\{\bm{f}_1,\ldots,\bm{f}_{|\calC|}\}$ stores candidate analog beams, and $\bF$ is constructed by selecting the beams that maximize received power, mutual information, signal-to-interference-plus-noise-ratio (SINR), or a scheduling metric.  DFT and hierarchical codebooks are natural for ULAs/UPAs and are easy to store on board. For LEO systems, a fixed terrestrial-style codebook may become inefficient
as the satellite footprint changes. Dynamic codebooks that adapt to the
orbital geometry can provide more stable beamforming gain over a satellite
pass~\cite{palacios_dynamic_2022,momani_beam_2025}. In UAV systems, location-aided beam selection can reduce training overhead by mapping UAV/user positions to a small candidate set before digital precoding~\cite{wu_location_2020,sugimoto_hybrid_2025}.

\subsubsection{Sparse recovery and OMP}
Because the channel in \eqref{eq:channel} is angularly sparse, HBF can be formulated as an approximation of a fully digital precoder by a sparse combination of array steering vectors.  A common form is expressed as
\begin{equation}
  \min_{\bF,\bB} \|\bF_{\mathrm{opt}}-\bF\bB\|_F^2
  \quad \text{s.t.} \quad |[\bF]_{i,j}|={1}/{\sqrt{\Nt}}, \ \forall i,j,
  \label{eq:fd_approx}
\end{equation}
where $\bF_{\mathrm{opt}}$ denotes the fully digital benchmark.  OMP greedily selects steering vectors from an overcomplete dictionary~\cite{hoang2025omp}, updates the residual, and solves a least-squares baseband problem.  This approach is instructive because it directly links HBF to channel sparsity, but its performance depends on the quality of the angular dictionary and can degrade under off-grid angles or wideband
beam squint~\cite{ahmed_survey_2018,khammassi_precoding_2023}.

\subsubsection{Manifold optimization}
The constant-modulus constraint defines a product of complex circles, which can be handled using Riemannian optimization.  In this approach, the Euclidean gradient of the objective is projected onto the tangent space of the manifold, followed by retraction back to the feasible set.  Manifold optimization is widely used in LEO and UAV HBF because it avoids exhaustive codebook search and can exploit continuous phase control
when high-resolution phase shifters are available~\cite{you_hybrid_2022,liu_robust_2022,zhang_uav-bs_2023}.  Its main limitation is that it provides local convergence and may require careful initialization.

\subsubsection{Wide-beam and robust analog designs}
Narrow analog beams maximize peak array gain but are sensitive to angular errors.  In LEO systems, angular errors arise from CSI aging, Doppler-induced prediction mismatch, and beam-squint effects.  In UAV systems, they are often caused by jitter and attitude
variations.  Robust analog designs intentionally broaden the beam or optimize the beam shape over an uncertainty interval.  Chirp sequence-inspired designs and adaptive beamwidth selection are representative UAV solutions~\cite{chen_adaptive_2024,liu_deployment_2023}, while statistical or geometry-aware designs have been explored for LEO and holographic terminals~\cite{li_holographic_2025}.  This family of methods is
important because NTN reliability is often limited by beam misalignment rather than by thermal noise alone.

\subsection{Digital Precoder Design and Optimization Frameworks}
\label{ssec:digital}

Once $\bF$ is fixed, the effective channel becomes $\tilde{\bH}=\bH\bF$, and the digital precoder $\bB$ operates in a reduced $\Nrf$-dimensional space.  This reduction is the main computational advantage of HBF, but it also means that the digital stage can only compensate for interference that remains after analog beam steering.

\subsubsection{ZF, RZF, and MMSE precoding}
ZF suppresses inter-user interference by projecting each user stream into the null space of the others.  For a full-row-rank effective channel, a representative ZF form is expressed as
\begin{equation}
  \bB_{\mathrm{ZF}} = \tilde{\bH}^{H}\big(\tilde{\bH}\tilde{\bH}^{H}\big)^{-1},
\end{equation}
followed by power normalization.  RZF and MMSE precoding improve robustness when the effective channel is ill-conditioned or the signal-to-noise ratio (SNR) is moderate.  These low-complexity precoders are common in satellite and UAV studies because they are simple enough for real-time implementation and can be combined with scheduling, clustering, or trajectory optimization~\cite{arnau_performance_2012,momani_beam_2025,wang_cooperative_2023}.

\subsubsection{WMMSE and fractional programming}
Weighted sum-rate maximization is non-convex in $\bB$. 
WMMSE converts this problem into an equivalent weighted mean-square-error minimization and solves the resulting subproblems iteratively.  
FP is similarly useful for energy-efficiency objectives, which are ratios of rate to power. 
These tools are especially important in NTN HBF because the objective is often not merely spectral efficiency: satellite studies include payload power, nonlinear amplifier distortion, and multicast constraints, while UAV studies include propulsion energy, battery endurance, and quality-of-service (QoS) requirements~\cite{you_hybrid_2022,liu_joint_2022,zhai_hybrid_2021,miao_serving_2025}.

\subsubsection{AO, SCA, SDP, and penalty-based decomposition}
The joint HBF problem is usually solved by alternating between $\bF$ and $\bB$.
AO is popular because the analog and digital subproblems have very different structures. 
Non-convex constraints are then handled by standard tools: SCA linearizes non-convex terms around the current iterate; semidefinite programming (SDP) relaxes rank-one matrix constraints; and penalty or augmented-Lagrangian methods enforce coupling constraints between variables.  
These frameworks appear across secrecy-energy-efficient satellite-terrestrial HBF, robust LEO HBF, RIS-assisted UAV HBF, and ISAC-oriented NTN HBF~\cite{lin_secrecy-energy_2021,liu_robust_2022,dong_hybrid_2023,wu_joint_2025,liu_hybrid_2025}.

\subsection{Machine Learning and Data-Driven HBF}
\label{ssec:ml}

Classical optimization algorithms are useful for benchmarking and offline design, but many of them require repeated matrix inversions, eigenvalue decompositions, or iterative convex solvers.  
This is problematic for NTN platforms where decisions must be made under limited onboard computation and rapidly changing geometry. Learning-based HBF therefore aims to shift computational burden to offline training and reduce online inference latency.

\subsubsection{Supervised deep learning}
Supervised DNNs can learn the mapping from CSI, location, or beam measurements to the hybrid precoder.  
In LEO systems, channel prediction is particularly attractive because orbital motion is deterministic and uplink/downlink channel correlation can be exploited to reduce feedback latency~\cite{Zhang2022LEOChannelPrediction,hong_deep_nodate}.  In UAV systems, supervised or cross-entropy-based learning can learn robust beam selection or hybrid precoding policies from simulated channel data~\cite{ren_machine_2019,wang_deep_2025}.

\subsubsection{Deep reinforcement learning}
DRL is more suitable when HBF is coupled with sequential control decisions.  
Examples include beam hopping in LEO satellites, where the scheduler must allocate beams over time, and UAV trajectory optimization, where movement changes both the channel and the energy budget.  
MAPPO and soft actor-critic (SAC)-type algorithms have been used for joint beamforming, power allocation, beam hopping, RIS control, or multi-UAV coordination~\cite{meng_joint_2025,hevesli_hybrid_2025,kakati_hybrid_2025,silvirianti_sub-connected_2023}.

\subsubsection{Model-driven learning}
A promising middle ground is model-driven learning, where known algorithmic steps are embedded into a trainable architecture.  
Instead of treating HBF as a black-box mapping, model-driven networks can unfold AO, OMP, or channel-estimation iterations.  
This reduces training complexity and improves interpretability, which is important for reliability-critical satellite and UAV applications.
Autoencoder-based pilot design and compressive beamspace channel estimation for LEO links illustrate this direction~\cite{ku_leveraging_2026}.

It is worth noting that learning-based HBF should not be interpreted as a replacement for physical modeling in NTN systems.  
Its strongest role is often to accelerate, approximate, or warm-start a model-based pipeline under predictable geometry and sparse channels.  
This is why hybrid model-based/learning-based designs are likely to be more deployable than purely black-box policies.

\subsection{Cross-Platform Design Lessons}
\label{ssec:taxonomy_lessons}

Sections~\ref{ssec:sysmodel}--\ref{ssec:ml} have filled in two of the four pillars of Fig.~\ref{fig:taxonomy_hbf_ntn}: the architecture and the methodology, which Table~\ref{tab:hbf_architecture_techniques} pairs against each other. The two remaining axes, platform and functional extension, are what Sections~\ref{sec:leo} and~\ref{sec:uav} develop. 
What this section already permits, however, is a set of design principles that hold across both platforms and that the two platform sections will substantiate.


This taxonomy leads to several cross-platform lessons.  
First, HBF should be viewed as an architecture--algorithm co-design problem.  
A mathematically strong precoder may be unsuitable if it requires an RF architecture that violates payload or battery limits.
Second, CSI quality is more important than CSI quantity.  
NTN channels are sparse and geometry-driven, so location, ephemeris, sensing, and statistical CSI can sometimes be more useful than delayed instantaneous CSI.  
Third, LEO and UAV systems require different notions of robustness, as Sections~\ref{sec:leo} and~\ref{sec:uav} will show in detail.  
LEO robustness is mainly against CSI aging, Doppler, beam squint, and long feedback delay, whereas UAV robustness is mainly against jitter, trajectory variation, and blockage. 
Finally, learning-based HBF is most credible when it is tied to physical structure, such as predictable orbital motion, beamspace sparsity, or trajectory-dependent channel evolution.

\section{HBF for LEO Satellite Systems}
\label{sec:leo}
{\color{black}To overcome the severe SWaP constraints that render fully digital massive MIMO impractical for LEO payloads, HBF maps high-dimensional antenna arrays to fewer RF chains via joint digital baseband and analog phase-shifting networks.}
To show how HBF adapts to the constraints of the space environment, this section reviews the literature along five categories. We first examine system architectures and hardware-constrained precoding designs. We then examine the coupled spatio-temporal dynamics that govern beam hopping. Next, we discuss the network-level dimension, which comprises multi-satellite cooperation, user scheduling, and CSI acquisition. We subsequently review the multi-functional physical layer, in which environmental sensing and reconfigurable surfaces are integrated into the HBF design. Finally, we review security, multiple access, and interference mitigation for wide-footprint satellite downlinks. Section~\ref{sec:uav} adopts the same five categories for UAV systems, so that the two platforms can be compared category by category.

\subsection{{\color{black}System Architecture and Precoding Design}}
\label{ssec:b1}
The design of hybrid precoding architectures for LEO satellite systems represents the most extensively studied topic in this section, addressing spectral or energy efficiency under stringent spaceborne hardware constraints. Unlike terrestrial BSs, LEO payloads operate under tight SWaP constraints, making the choice of HBF architecture a critical design decision. 
This subsection categorizes the literature into four groups, namely energy-efficient precoding under statistical CSI, robust and low-complexity precoding under mobility and CSI uncertainty, ground-terminal transceiver and prototyping architectures, and cell-free and wideband high-frequency architectures.

\subsubsection{{Energy-Efficient Precoding under Statistical CSI}}
The primary driver for adopting HBF in LEO constellations is the need to balance the high spatial multiplexing gain of massive MIMO with the power limitations of space-grade components. To address this, the hybrid analog/digital precoding framework derived in \cite{you_hybrid_2022} introduces a methodology that utilizes statistical CSI at the transmitter for the satellite downlink. Their work characterizes performance across both continuous and discrete phase-shift networks under fully and partially connected sub-array architectures. Recognizing the practical severity of hardware non-linearities in space environments, the authors subsequently extend their framework in \cite{you_massive_2022} to incorporate twin-resolution phase shifting (TRPS) and non-linear power amplifier (NPA) distortion. This extension provides a critical design framework for mitigating signal degradation caused by high-power downlinks operating near amplifier saturation. From a deployment and standardization perspective, the authors in \cite{palacios_hybrid_2021} shift the focus toward practical interoperability by developing a 5G New Radio (NR)-compatible HBF design, specifically optimizing analog codebooks for Ku-band LEO-to-user equipment (UE) links to ensure seamless integration with terrestrial cellular ecosystems.

\subsubsection{{Robust and Low-Complexity Precoding under Mobility and CSI Uncertainty}}
The high orbital velocity, i.e., ${\approx 7.5\text{ km/s}}$, of LEO satellites introduces severe Doppler shifts and propagation delays, rendering conventional, perfect-CSI precoding obsolete. To tackle the resulting channel aging and estimation errors, a robust energy-efficient HBF scheme is formulated in \cite{liu_robust_2022}, where the non-convex design problem is reformulated through SDP and solved by a nested quadratic transform fractional programming and constrained concave-convex procedure (QTFP-CCCP) algorithm.

Aside from channel uncertainty, the computational overhead of updating high-dimensional precoding matrices at millisecond intervals presents an on-board bottleneck for satellite processors. To mitigate this overhead, the framework in \cite{hwang_low-complexity_2025} leverages the high correlation between adjacent-frequency precoding vectors that naturally occurs in single-path satellite channels. Their compression-based hybrid precoding (CHP) method dramatically reduces computational complexity to just $15.5\%$ of baseline conventional methods without sacrificing sum-rate performance. Crucially, the practical operational rates of these architectures are quantified by \cite{momani_beam_2025}. Their work analyzes the update-rate requirements for analog beam steering versus digital precoding across various LEO altitudes, demonstrating a stark multi-scale temporal decoupling, i.e., analog phase-shifter coefficients only require updates on the order of seconds to track geometric trajectory shifts, whereas digital precoding coefficients must be updated at millisecond-level intervals to counter fast-fading and multi-user interference.

\subsubsection{{Ground-Terminal Transceiver and Prototyping Architectures}}
While the majority of literature targets the satellite payload, the complementary problem of HBF design at the ground station (GS) or user terminal is equally vital, particularly when transitioning to high-frequency bands. A geometry-based precoding algorithm for mmWave ground-station HBF is validated on a hardware-in-the-loop software-defined radio testbed in \cite{struyf_hybrid_2025}. On the hardware manufacturing front, the research in \cite{chi_hybrid_2023} designs and empirically evaluates a hybrid multibeamforming receiver for Ku-band LEO communications. Their transceiver achieves ultra-fine phase-control resolutions of $0.022^\circ$ in the analog domain and $0.72^\circ$ in the digital domain, mapping out the precise hardware bounds for modern satellite multi-beam synthesis. These empirical designs build upon early foundational theory established in \cite{ivanov_spatial_2020, ivanov_physical_2019}. Their seminal works on spatial resource management and physical-layer HBF structures provide the initial mathematical justification for selecting hybrid architectures over fully analog or fully digital baselines under realistic constraints of power consumption and non-uniform user distributions.

\subsubsection{{Cell-Free and Wideband High-Frequency Architectures}}
As satellite networks evolve toward 6G, HBF is increasingly paired with next-generation networking topologies and high-frequency spectrum frontiers. A network topology merging user-centric cell-free massive MIMO deployment models with dense LEO satellite networks is introduced in \cite{miao_serving_2025}. They propose a cooperative access point (AP) clustering strategy alongside a distortion-aware hybrid precoder designed to suppress NPA-induced signal degradation across co-serving satellites. To maximize spectral utilization within the satellite payload itself, the authors in \cite{hsiao_digital_2024} examine partially spectral-overlapping subarrays, deriving an MMSE-based digital precoder that significantly improves spectral efficiency while robustly mitigating inter-beam interference.

Furthermore, the expansion into the THz band to achieve massive data rates introduces the destructive phenomenon of beam squint. The work \cite{wu_hybrid_2023} addresses this by combining HBF with an ITS, proposing an angle-based hybrid beamformer that effectively suppresses beam squint across wideband non-terrestrial links. Finally, to handle the core non-convex constraints of phase-shifter quantization in single-satellite configurations, the authors in \cite{kong_hybrid_nodate} develop an energy-efficient HBF framework utilizing a combination of AO and manifold optimization, offering a mathematically rigorous approach to bounding the performance loss of low-resolution hardware.

\subsection{{Beam Hopping and Spatio-Temporal Beam Management}}
\label{ssec:b2}
Beam hopping (BH) is a critical, multibeam satellite-specific technique wherein available time slots are dynamically allocated to a subset of spot beams in response to non-uniform, time-varying traffic demands. This operational paradigm enables highly flexible resource allocation without incurring the prohibitive payload hardware costs and power consumption associated with illuminating all beams simultaneously. However, the interplay between BH and HBF introduces a deeply coupled spatio-temporal optimization problem that has no direct analogue in cellular or low-altitude UAV networks. In this architecture, the analog beamformer dictates the spatial steering vectors across active time slots, while the digital precoder manages instantaneous intra-beam interference and power distribution. Consequently, the joint scheduling of time-domain beam illumination patterns and space-domain beamforming vectors yields highly non-convex, mixed-integer non-linear programming (MINLP) problems that must be solved under strict latency constraints.

\subsubsection{{Dynamic and Multi-Satellite Beam-Hopping Scheduling}}
To address the computational complexity of real-time resource allocation, recent literature heavily leverages artificial intelligence and advanced optimization theory. To accommodate highly dynamic traffic profiles, the framework in \cite{meng_joint_2025} addresses the joint optimization of HBF alongside real-time beam illumination patterns. They propose an MAPPO framework that elegantly decomposes the problem, i.e., the deep reinforcement learning agents dictate the time-slot power allocation, while ZF principles are applied to solve the digital precoding matrix in the spatial domain. Scaling this problem from a single payload to a network-wide deployment, the authors in \cite{cui_joint_2025} extend the BH paradigm to dense multi-LEO-satellite systems. To manage the massive spatial degrees of freedom and inter-satellite interference, they propose a dual-layered matching theory framework in which a many-to-one algorithm handles ground-user-to-satellite cell association, while a many-to-many matching algorithm determines the optimal BH illumination patterns. This scheduling layer is tightly coupled with a physical-layer quadratic transform-based precoding design, maximizing system throughput across the constellation.

\subsubsection{{\color{black}Illumination-Pattern Synthesis and Matrix Optimization}}
Analog phase-shifter constraints introduce severe combinatorial bottlenecks into BH matrix optimization. To address this layout under hybrid precoding constraints, the works in \cite{han_beam_2024, han_beam_2023} formulate a joint scheduling optimization aimed at minimizing total transmit power while guaranteeing strict user rate requirements. They examine both structured and irregular BH matrices, proposing a Viterbi-based optimization for regular matrices and a low-complexity greedy selection algorithm for irregular variants. Their findings indicate that the Viterbi approach leverages the temporal memory of the scheduling state to outperform the greedy baseline by up to $2.8\text{ dB}$ in power efficiency. The inherent structural complexity is successfully decoupled in \cite{wang_hybrid_2023} through a sequential optimization strategy rooted in fractional programming. Their framework first determines the optimal BH illumination patterns under an idealized, fully digital beamforming assumption, establishing a performance upper bound. They subsequently map these idealized vectors onto practical, constrained hybrid precoding structures using alternating minimization, minimizing the quantization loss inherent in the analog weight network.

\subsubsection{{RF Front-End Dynamics and Predistortion under Fast Beam Switching}}
\begin{figure*}[t]
	\centering
	\includegraphics[width=.95\textwidth]{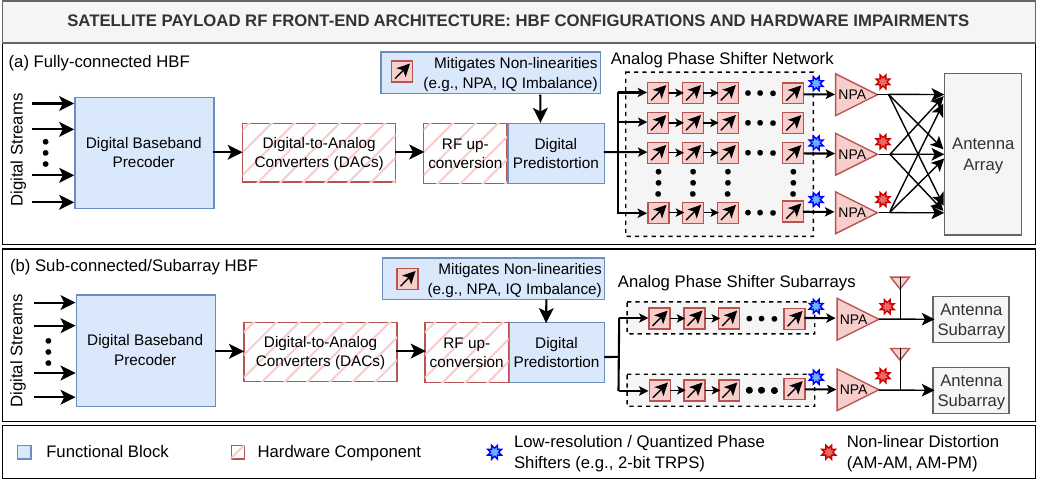}
	\caption{Structural comparison of satellite payload HBF architectures under hardware impairments. a) Fully-Connected HBF structure, mapping every RF chain to all antenna elements via a low-resolution analog phase-shifter network. b) Sub-Connected / Subarray HBF structure, optimizing power consumption via dedicated localized phase-shifter subarrays. Both configurations illustrate the integration of DPD frameworks to mitigate NPA distortions and TRPS quantization errors right before the radiating antenna aperture~\cite{you_massive_2022, deng_real-time_2025}.}
	\label{fig:HBF_HI}
    \vspace{-0.25cm}
\end{figure*}

A highly critical, yet frequently overlooked, challenge in high-speed BH is the behavior of the satellite's RF front-end during microsecond- or millisecond-scale beam transitions. Rapidly toggling analog beam weights causes severe transient variations in the load impedance of the on-board high-power amplifiers (HPAs), exacerbating their inherently non-linear characteristics. To bridge the gap between abstract scheduling theory and physical implementation, the authors in \cite{deng_real-time_2025} propose a real-time spatial digital predistortion (DPD) framework specifically tailored for HBF-BH architectures. Their solution introduces a power-indexed DPD codebook that successfully decouples the high-speed DPD linearization coefficients from the macro-level BH scheduling loop. This hardware-aware design ensures that aggressive spatial nulling and beam steering remain stable and free of spectral regrowth, even during rapid, millisecond-scale illumination handovers. Fig.~\ref{fig:HBF_HI} maps the detailed schematic of both fully-connected and subarray hybrid precoding architectures under realistic hardware distortions.

\subsection{{Network-Level Design: Multi-Satellite Cooperation, Scheduling, and CSI Acquisition}}
\label{ssec:b3}
This category groups three interconnected topics that collectively address the \emph{network} dimension of satellite HBF, which consists of how multiple spaceborne nodes cooperate to form distributed networks, how ground users are scheduled across highly mobile footprints, and how CSI is acquired, tracked, or predicted despite severe propagation delays and Doppler shifts. In single-satellite systems, HBF performance is limited by the individual payload’s field of view. Transitioning HBF to the network level requires coordinating spatial resources across overlapping satellite footprints while simultaneously managing severe channel aging caused by rapid orbital mobility.

\subsubsection{{Multi-Satellite Cooperation and Distributed Space-MIMO}}
When multiple LEO satellites illuminate the same geographic area, inter-satellite interference can severely degrade user SINRs. Conversely, if tightly coordinated, these overlapping beams can be exploited as a distributed spaceborne MIMO transmitter. 
The authors in \cite{zhang_multi-satellite_2023, zhang_joint_2023} pioneer the study of HBF for cooperative multi-LEO networks. Their architecture decouples
the beamforming process into two distinct tasks, including analog vectors that handle macroscopic beam alignment to track the moving satellites, while digital precoders are dynamically optimized to eliminate multi-user and inter-satellite interference. When coupled with their heuristic user scheduling algorithm based on spectral efficiency increments, this joint scheme achieves a remarkable $47.2\%$ spectral efficiency gain over conventional non-cooperative baselines.

To expand the physical-layer aperture, the framework in \cite{xu_distributed_2025} models a cluster of LEO satellites as a unified, distributed space-MIMO transmitter. They propose a graph-based distributed HBF scheme tailored for RHS-empowered links, providing a scalable optimization framework that they subsequently extend to traditional phased-array configurations to validate its architectural flexibility.

\subsubsection{{User Scheduling and Resource Allocation}}
Because the number of ground users in a satellite cell typically far exceeds the available RF chains on the payload, intelligent user scheduling is mandatory to maximize the multi-user multiplexing capability of digital precoders. Within the context of the DVB-S2X standard framework, a robust joint user scheduling and HBF design is developed in \cite{liu_joint_2022} for massive MIMO LEO multigroup multicast systems. Their design explicitly accounts for residual Doppler shifts caused by imperfect synchronization, utilizing a hierarchical clustering algorithm to group users with highly correlated channel matrices before executing a semidefinite programming constrained concave-convex procedure (SDP-CCCP) optimization.

In the context of the integrated satellite-terrestrial network (ISTN), where spectrum is aggressively shared between space and cellular infrastructure, the authors in \cite{deyi_hybrid_2021} develop an adaptive user scheduling scheme based on dynamic channel correlation thresholds. The resulting non-convex power allocation problem is resolved using MMSE criteria combined with logarithmic linearization techniques. To minimize total transmit power in multi-user massive MIMO-OFDM LEO downlinks, a singular value decomposition (SVD)-based user selection framework is introduced in \cite{wang_user_2024}. By executing this selection independently across individual frequency subchannels, the algorithm precisely maps the ground user distribution to the dominant spatial eigenmodes of the satellite channel.

\subsubsection{{CSI Acquisition, Prediction, and Limited Feedback}}
The efficacy of any HBF framework hinges on the availability of accurate CSI. However, for LEO systems, the round-trip propagation delay can exceed the channel coherence time, resulting in severe channel aging. To bypass the traditional feedback-delay bottleneck, a deep-learning-driven framework is proposed in \cite{Zhang2022LEOChannelPrediction} comprising two neural networks, termed SatCP and SatHB. This architecture leverages a deep neural network to predict future downlink CSI directly from observed uplink channel metrics by exploiting inherent uplink-downlink spatial correlations. The hybrid beamformer is then synthesized immediately from the predicted CSI, entirely bypassing iterative mathematical optimization loops.

Where explicit feedback is required, the design of the analog codebook dictates the stability of the tracking link. The work in \cite{palacios_dynamic_2022} reflects a dynamic codebook design for analog beamforming that shifts its phase profiles in accordance with the satellite's deterministic trajectory. This approach provides a highly stable beamforming gain and significantly higher SINR compared to static DFT codebooks as the satellite sweeps across its coverage zone.
To bypass heavy iterative CSI feedback under severe channel aging, a limited-feedback HBF scheme is studied in \cite{11575120} for multiuser LEO downlinks under beam misalignment and SINR aging. The framework pairs codebook-based beam-index selection with low-rate, quantized SINR reports to drive a fairness-oriented power allocation. By maximizing a soft-min utility via a low-dimensional simultaneous perturbation stochastic approximation algorithm, this design robustly protects weak users. Sensitivity analyses show that despite highly constrained feedback granularity, this approach increases 5th-percentile and per-drop minimum rates by up to 20--30\% over SINR-greedy benchmarks, balancing overall sum-rate with feedback-efficient user fairness.

Finally, to address the pilot overhead bottleneck in high-dimensional beamspace channels, a joint auto-encoder-based framework for pilot waveform design and compressive sensing-based channel estimation is proposed in \cite{ku_leveraging_2026}. By demonstrating that 1-sparse training patterns enable efficient OMP-based decoding at the receiver, their framework successfully recovers full beamspace CSI using significantly fewer pilot time slots, preserving vital time-frequency resources for data transmission.

\subsection{{Sensing Capability and Reconfigurable Surfaces}}
\label{ssec:b4}
A growing body of literature integrates environmental sensing capabilities and reconfigurable electromagnetic surfaces into the satellite HBF framework, introducing a multi-functional physical layer that extends well beyond traditional data communication. By unifying the hardware infrastructure required for radar sensing, target localization, and spatial signal reflection, these emerging architectures maximize the utility of spaceborne payloads. This subsection reviews recent advancements across two core paradigms consisting of ISAC and RIS, or holographic metasurface-aided satellite links.

\subsubsection{{\color{black}Integrated Sensing and Communication}}
Integrating radar sensing and communication into a unified LEO hardware platform allows satellites to simultaneously serve ground-based communication terminals while tracking space debris, low-altitude aircraft, or geographic features. The authors in \cite{you_integrated_2024} investigate joint communication and localization frameworks for massive MIMO LEO systems, demonstrating that mathematically optimized hybrid precoding matrices can simultaneously support high-throughput downlink data transmission and high-precision user positioning.

To address the physical constraints of wideband mmWave transmission over large-scale satellite antenna arrays, the same group \cite{you_beam_2022} derives a beam squint-aware ISAC design for LEO payloads. In wideband regimes, frequency-dependent spatial phase shifts across the subcarriers cause the physical beam to drift a phenomenon, known as beam squint. To counter this, the authors develop a specialized wideband hybrid precoding architecture that unifies spatial focusing across the entire OFDM spectrum, thereby maintaining an optimal trade-off between sharp radar tracking and high-capacity communication.

\begin{figure}[t]
	\centering
	\includegraphics[width=\linewidth]{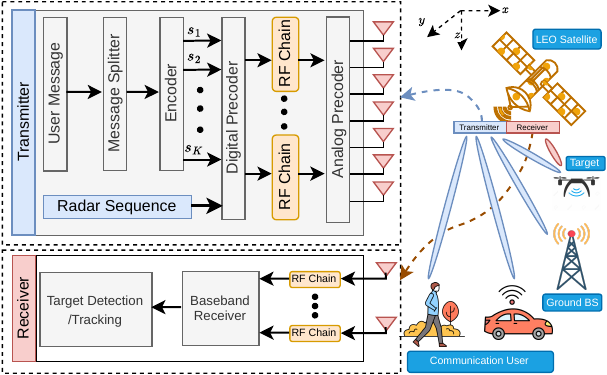}
	\caption{Functional block diagram of an ISAC-enabled HBF LEO satellite downlink. The schematic maps the joint processing loop where communication data and radar probing waveforms are synthesized via a unified digital baseband precoder and an analog phase-shifting network. On the ground, the communication user isolates and decodes its designated data stream from the communication beams, while the reflected radar echo from the sensing target is captured and processed by the satellite receiver for high-precision target detection and localization \cite{you_integrated_2024, cao_robust_2025}.}
	\label{fig:NTN_ISAC_signal}
    \vspace{-0.25cm}
\end{figure}

Moving toward complex multi-user access networks, a partially connected HBF design for a rate-splitting multiple access (RSMA)-enabled LEO-ISAC system is introduced in \cite{liu_hybrid_2025} tailored for the power Internet of Things (IoT). To overcome the severe full-duplex self-interference inherent when a single payload transmits and receives radar chirps simultaneously, they introduce a dual-array architecture utilizing a spatially separated assisting receiver. They deploy a PDD algorithm to solve the highly coupled multi-target sensing and interference-mitigation problem across two distinct operational cases, verifying the robustness of RSMA under hardware-constrained HBF-ISAC nodes. The signal-flow architecture of an ISAC-enabled HBF LEO satellite downlink is illustrated in Fig.~\ref{fig:NTN_ISAC_signal}.

\subsubsection{{RIS-Assisted and Holographic Metasurface Transceivers}}
Deploying RISs and RHS offers a low-cost, energy-efficient method to bypass LoS blockages in dense urban or mountainous terrain and to synthesize highly directional beams at the user terminal (UT) side. To address multi-altitude LEO network topologies, the work in \cite{li_holographic_2025} develops a specialized HBF architecture for holographic metasurface-based user terminals. To mitigate the computational overhead of tracking instantaneous channel variations across highly dynamic multi-altitude orbits, a low-complexity MMSE algorithm is developed grounded in stochastic geometry that operates entirely on statistical CSI.

At the transceiver hardware level, the work in \cite{deng_holographic_2022} introduces RHS technology explicitly for LEO satellite user terminals. By modeling the continuous aperture illumination of holographic surfaces, the authors derive a closed-form optimal holographic beamformer for sum-rate maximization. Their empirical performance bounds demonstrate that RHS structures achieve superior beam-steering resolution and higher sum-rates than traditional discrete phased arrays occupying the same physical aperture, offering a promising low-SWaP alternative for mobile ground stations.

Finally, the geographic positioning of these smart surfaces dictates their ultimate beamforming efficiency. The proposed methodology in \cite{kang_leo_2024} focuses on optimizing the physical deployment location of a ground-based RIS within an LEO HBF downlink framework. Their spatial optimization models reveal that strategic, geometry-aware RIS placement can compensate for the hardware degradation of sub-connected analog phase-shifter networks, driving the overall system sum-rate performance remarkably close to that of an unconstrained, fully digital beamforming architecture.

\subsection{{Security, Multiple Access, and Interference Mitigation}}
\label{ssec:b5}
Because satellite transmissions span immense geographic footprints, their broadcast signals are inherently vulnerable to unauthorized interception, malicious jamming, and unintentional spectral overlap with terrestrial networks. Resolving these challenges requires HBF designs that integrate multi-user isolation, secure directional steering, and cross-network interference suppression. This subsection reviews recent literature addressing these multi-dimensional operational threats through three lenses including physical layer security (PLS), advanced multiple access schemes such as RSMA and non-orthogonal multiple access (NOMA), and spectrum coexistence paradigms.

\subsubsection{{Physical Layer Security}}
Due to the wide coverage area of LEO spot beams, legitimate user signals often spill over into regions containing malicious or unauthorized eavesdroppers. Physical layer security exploits the spatial degrees of freedom provided by HBF to steer directional nulls toward eavesdroppers while maintaining high directive gains toward legitimate terminals. The secrecy-energy-efficient HBF within an ISTN network is investigated in \cite{lin_secrecy-energy_2021}. In their model, a multi-beam LEO satellite shares mmWave frequencies with a terrestrial cellular network under the threat of active eavesdropping. Recognizing that precise interceptor coordinates are rarely available in tactical spaceborne scenarios, the authors model imperfect wiretap channel angles. They derive two robust beamformer designs by leveraging the Charnes-Cooper transformation paired with SCA, effectively maximizing the secrecy-to-power consumption ratio under bounded channel uncertainty.

\subsubsection{{Rate-Splitting and Non-Orthogonal Multiple Access}}
To handle dense user distributions, conventional orthogonal multiple access (OMA) is increasingly replaced by non-orthogonal multiplexing strategies tailored for HBF frameworks. RSMA has emerged as a particularly robust strategy for LEO architectures because it gracefully degrades under the imperfect CSI at the transmitter caused by satellite mobility. The authors in \cite{cao_robust_2025} integrate RSMA with a wide-beam analog precoder configuration to design a robust HBF scheme for multi-beam LEO satellite downlinks. Their framework explicitly accounts for spatial alignment angle errors and imperfect CSI, demonstrating that coupling the common-stream and private-stream mechanics of RSMA with analog phase-shifter constraints provides superior robustness against channel aging compared to conventional ZF HBF.

In parallel, NOMA has been adapted to LEO payloads to increase the user-carrying capacity per beam. The work in \cite{li_maximizing_2023} applies NOMA to an HBF-equipped LEO satellite system, focusing on the highly coupled problem of user grouping and power allocation. They utilize the agglomerative nesting (AGNES) hierarchical clustering algorithm to group users exhibiting high channel spatial correlation into shared beam resources. They subsequently optimize the network's overall energy efficiency by executing FP and SCA, demonstrating that NOMA-HBF architectures significantly outperform OMA-HBF benchmarks when power resources are severely constrained.

\subsubsection{{Interference Mitigation and Spectrum Coexistence}}
The aggressive rollout of mega-constellations creates severe coexistence bottlenecks, particularly when satellites share spectrum with incumbent terrestrial infrastructure. The authors in \cite{razavi_interference-aware_2025} address this problem by investigating the co-channel coexistence of terrestrial 5G BSs (gNBs) and LEO satellite downlinks within the upper mid-band spectrum in $7\text{--}24\text{ GHz}$. They propose an interference-aware hybrid precoding algorithm that integrates a dynamic LEO protection penalty directly into the optimization objective function. Their hardware-constrained design establishes sharp spatial notches aimed at the angular directions of terrestrial assets. Empirical evaluations reveal that this protection framework dramatically reduces the probability of violating strict satellite-to-ground interference thresholds while preserving the overall satellite network sum-rate to within $3\%$ of unconstrained HBF baselines.

\subsection{Summary and Lessons Learned}

\begin{table*}[t]
\centering
\caption{\normalfont\scshape Summary of Representative HBF Architectures and Design Techniques for LEO Satellite Systems}
\label{tab:leo_survey_comparison}
\renewcommand{\arraystretch}{1.15}
\setlength{\tabcolsep}{2.6pt}
\scriptsize
\begin{tabular}{|c|c|
>{\raggedright\arraybackslash}p{3.30cm}|l|l|
>{\raggedright\arraybackslash}p{2.85cm}|}
\hline
\cellcolor{myblue}{\textbf{Reference}} &
\cellcolor{myblue}{\textbf{Architecture}} &
\multicolumn{1}{c|}{\cellcolor{myblue}{\textbf{Methodology}}} &
\multicolumn{1}{c|}{\cellcolor{myblue}{\textbf{Platform Coupling}}} &
\multicolumn{1}{c|}{\cellcolor{myblue}{\textbf{Key Enabling Technique}}} &
\multicolumn{1}{c|}{\cellcolor{myblue}{\textbf{Objective}}} \\
\hline\hline
\multicolumn{6}{|l|}{\cellcolor{gray!15}\textit{System Architecture and Precoding Design}} \\
\hline
\cite{you_hybrid_2022} & FC/PC & Statistical-CSI precoding & Statistical CSI & Continuous versus discrete phase-shift networks & SR \\
\hline
\cite{you_massive_2022} & FC/PC & Statistical-CSI precoding & Statistical CSI & TRPS quantization + NPA distortion model & SR \\
\hline
\cite{palacios_hybrid_2021} & FC & Codebook design & NR interoperability & Ku-band NR-compatible analog codebook & SR \\
\hline
\cite{liu_robust_2022} & FC & SDP + QTFP-CCCP & Doppler, channel aging & Robust design under bounded CSI error & EE \\
\hline
\cite{hwang_low-complexity_2025} & FC & Compression-based precoding & Single-path channel & Adjacent-subcarrier precoder correlation & SR at 15.5\% complexity \\
\hline
\cite{momani_beam_2025} & FC & Update-rate analysis & Orbital altitude & Analog/digital timescale decoupling & Tracking update rate \\
\hline
\cite{struyf_hybrid_2025} & FC & Geometry-based precoding & Ground station & Hardware-in-the-loop testbed validation & Link validation \\
\hline
\cite{chi_hybrid_2023} & FC & Transceiver hardware design & Ground terminal & Ku-band hybrid multibeam receiver & Phase-control resolution \\
\hline
\cite{ivanov_spatial_2020,ivanov_physical_2019} & FC/PC & Analytical & Non-uniform user density & Spatial resource management foundations & Power / SR trade-off \\
\hline
\cite{miao_serving_2025} & Cell-free & AP clustering & Multi-satellite & Distortion-aware cooperative precoding & SR \\
\hline
\cite{hsiao_digital_2024} & PC & MMSE & Payload spectrum reuse & Partially spectral-overlapping subarrays & SE \\
\hline
\cite{wu_hybrid_2023} & ITS-assisted & Angle-based design & Wideband THz & Beam-squint suppression & SE \\
\hline
\cite{kong_hybrid_nodate} & FC & AO + manifold optimization & Phase quantization & Low-resolution phase-shifter loss bounding & EE \\
\hline\hline
\multicolumn{6}{|l|}{\cellcolor{gray!15}\textit{Beam Hopping and Spatio-Temporal Beam Management}} \\
\hline
\cite{meng_joint_2025} & FC & MAPPO + ZF & Traffic dynamics & DRL power allocation with ZF digital stage & Traffic-matched SR \\
\hline
\cite{cui_joint_2025} & FC & Matching theory + FP & Multi-satellite & Two-layer user-cell and BH pattern matching & Throughput \\
\hline
\cite{han_beam_2024,han_beam_2023} & FC & Viterbi / greedy selection & BH scheduling & Structured versus irregular BH matrices & Power (2.8\,dB over greedy) \\
\hline
\cite{wang_hybrid_2023} & FC & FP + alternating minimization & BH scheduling & Digital upper bound mapped to hybrid stage & SR \\
\hline
\cite{deng_real-time_2025} & FC/PC & Power-indexed DPD codebook & BH transitions & DPD decoupled from the scheduling loop & Linearity under switching \\
\hline\hline
\multicolumn{6}{|l|}{\cellcolor{gray!15}\textit{Network-Level Design: Cooperation, Scheduling, and CSI}} \\
\hline
\cite{zhang_multi-satellite_2023,zhang_joint_2023} & FC & Two-stage + heuristic scheduling & Multi-satellite & Analog tracking with digital interference nulling & SE (+47.2\%) \\
\hline
\cite{xu_distributed_2025} & Holographic & Graph-based optimization & Satellite cluster & Distributed space-MIMO over RHS links & SR \\
\hline
\cite{liu_joint_2022} & FC & Clustering + SDP-CCCP & Residual Doppler & DVB-S2X multigroup multicast scheduling & Multicast rate \\
\hline
\cite{deyi_hybrid_2021} & FC & MMSE + log linearization & ISTN coexistence & Correlation-threshold user scheduling & Power \\
\hline
\cite{wang_user_2024} & FC & SVD & MIMO-OFDM downlink & Per-subchannel user selection & Power minimization \\
\hline
\cite{Zhang2022LEOChannelPrediction} & FC & DNN (SatCP, SatHB) & Channel aging & Uplink-downlink correlation for CSI prediction & SR without iteration \\
\hline
\cite{palacios_dynamic_2022} & FC & Codebook design & Orbital trajectory & Trajectory-shifted dynamic codebook & SINR \\
\hline
\cite{11575120} & FC & SPSA on soft-min utility & Beam misalignment & Beam index with quantized SINR feedback & Fair rate (+20--30\%) \\
\hline
\cite{ku_leveraging_2026} & Beamspace & Auto-encoder + OMP & Beamspace sparsity & Jointly designed pilot waveforms & Pilot overhead \\
\hline\hline
\multicolumn{6}{|l|}{\cellcolor{gray!15}\textit{Sensing Capability and Reconfigurable Surfaces}} \\
\hline
\cite{you_integrated_2024} & FC & Joint optimization & Statistical CSI & Shared precoder for data and localization & SR / positioning \\
\hline
\cite{you_beam_2022} & FC & Wideband hybrid precoding & Beam squint & Spatial focusing across the OFDM band & Sensing / SR trade-off \\
\hline
\cite{liu_hybrid_2025} & PC & PDD & Full-duplex self-interference & Dual array with separated assisting receiver & RSMA rate / sensing \\
\hline
\cite{li_holographic_2025} & Holographic & MMSE + stochastic geometry & Multi-altitude orbits & Statistical-CSI holographic user terminal & SR \\
\hline
\cite{deng_holographic_2022} & Holographic & Closed-form & User terminal SWaP & Continuous-aperture RHS beamformer & SR \\
\hline
\cite{kang_leo_2024} & RIS-assisted & Spatial optimization & RIS placement & Geometry-aware ground RIS siting & SR \\
\hline\hline
\multicolumn{6}{|l|}{\cellcolor{gray!15}\textit{Security, Multiple Access, and Interference Mitigation}} \\
\hline
\cite{lin_secrecy-energy_2021} & FC & Charnes-Cooper + SCA & ISTN coexistence & Robust nulling under wiretap angle error & Secrecy EE \\
\hline
\cite{cao_robust_2025} & FC & Robust optimization & Alignment angle error & RSMA common and private streams & SR (robust) \\
\hline
\cite{li_maximizing_2023} & FC & AGNES + FP + SCA & Power-limited payload & NOMA user grouping by spatial correlation & EE \\
\hline
\cite{razavi_interference-aware_2025} & FC & Penalty-based optimization & Terrestrial coexistence & Spatial notches toward incumbent gNBs & SR (within 3\% of baseline) \\
\hline
\end{tabular}
\vspace{-0.25cm}
\end{table*}
Table~\ref{tab:leo_survey_comparison} consolidates the surveyed LEO works along the same five columns later used for UAV systems in Table~\ref{tab:uav_survey_comparison}, so that the two platforms can be read against each other. Both tables abbreviate the recurring design objectives as sum rate (SR), spectral efficiency (SE), energy efficiency (EE), and max-min rate (MMR), the last being the fair-rate criterion that maximizes the worst user's throughput. The platform-coupling column is where the contrast is
sharpest: for LEO the coupling is to quantities the payload cannot change but can predict, namely orbital trajectory, Doppler, beam squint, and channel aging, whereas Table~\ref{tab:uav_survey_comparison} shows UAV designs coupling to quantities the platform itself controls, namely altitude, position, and trajectory. This is why statistical CSI, ephemeris-driven codebooks, and prediction recur here, while joint placement-beamforming optimization recurs there.

Based on this comprehensive literature mapping, several key technical insights and lessons learned emerge:
\begin{itemize}
	\item \textit{Multi-Scale Temporal Decoupling Essential for SWaP Bounds:} A critical hardware insight is that analog and digital precoding matrices operate on fundamentally decoupled time horizons due to LEO orbital geometry. While analog beam-steering coefficients remain stable on the order of seconds to track deterministic trajectory shifts, digital precoder weights require millisecond-level adaptations to counter multi-user interference and fast fading. Exploiting this multi-scale temporal property is vital to minimizing on-board computational overhead and preserving satellite power.
	
	\item \textit{Hardware-Aware Predistortion is Imperative for Dynamic Scheduling:} Aggressive spaceborne resource allocation strategies, such as microsecond-scale beam hopping, introduce severe transient non-linearities at the HPA front-end. Consequently, abstract scheduling designs must be co-designed with real-time, spatial DPD frameworks. Decoupling linearization codebooks from the primary macro-level scheduling loop represents a crucial design prerequisite for avoiding spectral regrowth during rapid beam handovers.
	
	\item \textit{Transition from Reactive to Predictive CSI Networks:} Due to massive round-trip propagation delays exceeding channel coherence intervals, conventional closed-loop CSI feedback loops inevitably suffer from severe channel aging bottlenecks. The literature demonstrates a clear paradigm shift toward predictive CSI acquisition. By leveraging statistical geometry or deep learning networks to estimate downlink channels directly from uplink observations, satellite payloads can entirely bypass iterative signaling overhead and latency.
	
	\item \textit{Architectural Robustness via Non-Orthogonal Signaling:} Under real-world constraints such as spatial beam squint in high-frequency bands, e.g., mmWave/THz, and angular alignment errors, conventional orthogonal beamforming degrades rapidly. Integrating flexible access schemes like RSMA provides an intrinsic layer of robustness. By absorbing spatial alignment errors into a wide-beam common stream, RSMA-HBF configurations degrade gracefully under imperfect CSI, making them highly suited for highly dynamic multi-altitude constellations.
\end{itemize}

\section{HBF for UAV Systems}
\label{sec:uav}


Unlike LEO satellites, whose mobility follows deterministic orbital mechanics and whose hardware is fixed once launched, UAVs offer controllable three-dimensional mobility. 
This controllability is, at the same time, the defining design challenge of UAV HBF: the wireless channel, the analog beam, and the platform itself are no longer independent quantities, since changing the UAV's altitude or trajectory directly reshapes the very channel that the beamformer is trying to serve.
UAV HBF research is further shaped by mechanical platform jitter, tight battery and payload budgets, and a flexible relay topology that has no direct counterpart in fixed terrestrial deployments or in orbital satellite systems. 
%
Accordingly, this section reviews UAV HBF along the same five categories used for LEO systems in Section~\ref{sec:leo}, so that the two platforms can be compared category by category, namely (\textit{i}) system architecture and precoding design, (\textit{ii}) mobility-aware beam tracking and management, (\textit{iii}) network-level design based on multi-UAV cooperation and relaying, (\textit{iv}) sensing capability and reconfigurable surfaces, and (\textit{v}) security and multiple access. 
Their platform-specific content differs most sharply in category (\textit{ii}), because the dominant time-varying mechanism is traffic-driven beam hopping on a satellite payload, whereas it is platform mobility and mechanical jitter on a UAV.

\subsection{\textcolor{black}{System Architecture and Precoding Design}}
\label{ssec:c1}

\begin{figure*}[!t]
\centering
{\includegraphics[width=.95\linewidth]{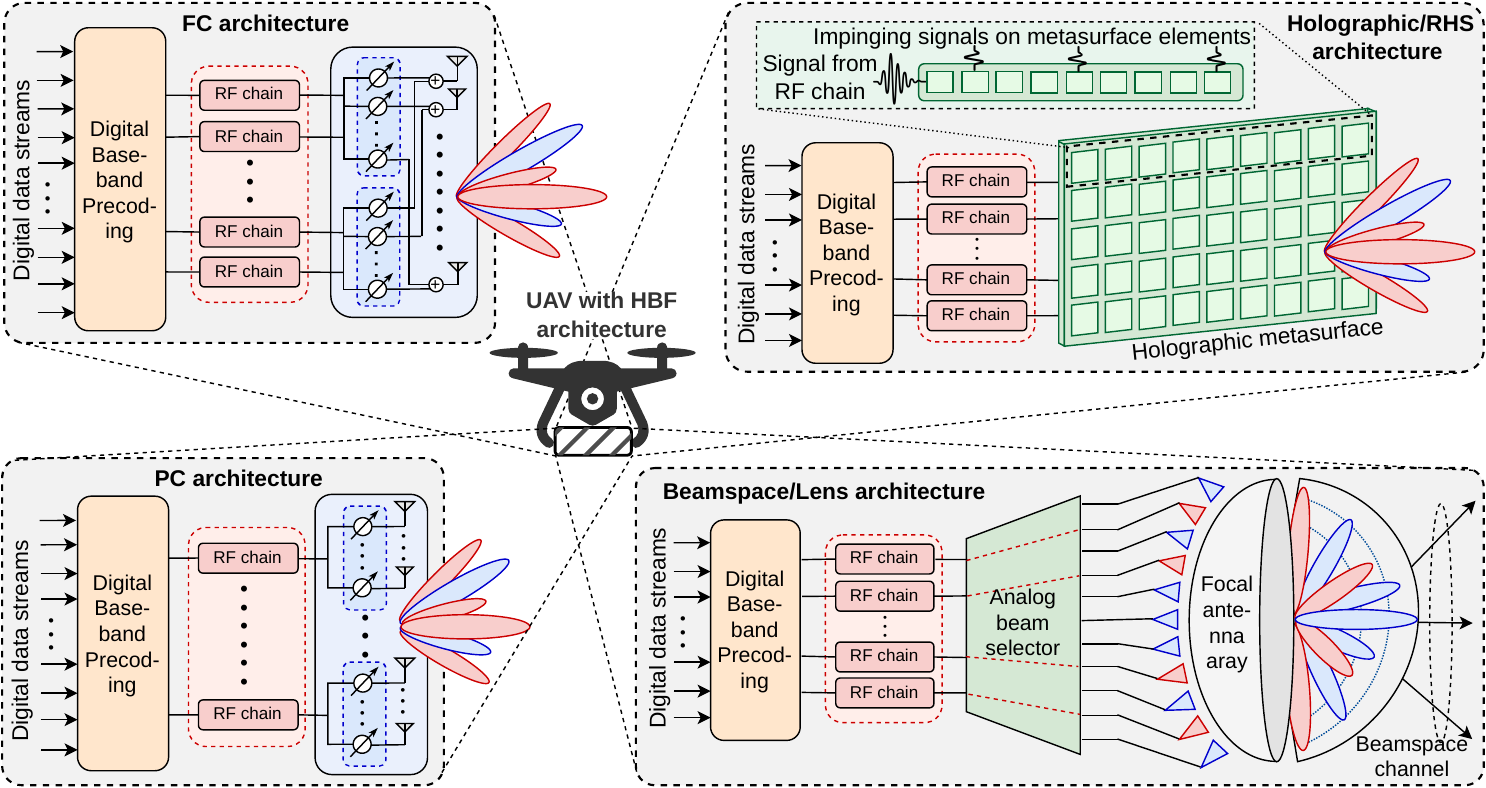}}
\caption{Four UAV-relevant HBF hardware architectures: 
(a) FC architecture: maximum analog flexibility, highest phase-shifter and power cost.
(b) PC architecture: each RF chain feeds a disjoint subarray, reducing hardware cost at the expense of beamforming freedom.
(c) Holographic/RHS architecture: a thin reconfigurable metasurface replaces the phased array, favoring the low-weight, planar UAV payload form factor.
(d) Beamspace/Lens architecture: a passive lens maps angular sparsity onto a small number of RF chains.
}
\label{fig:uav_architecture}
\vspace{-0.25cm}
\end{figure*}

{A UAV BS can employ the same FC and PC architectures as a terrestrial transmitter, as introduced in Section~\ref{ssec:sysmodel}. However, the selection between the two architectures is not governed by spectral efficiency alone. The reason is that the mass and power budget of a small or medium rotary-wing UAV is much lower than that of a terrestrial BS, so that the number of phase shifters, the RF front-end power, and the digital baseband load compete directly with the battery capacity and hence with the flight time of the platform. In addition, the altitude and the lateral position of the UAV determine the angular spread and the elevation angle of the channel toward the served ground users. Therefore, the array geometry, the number of antennas, and the platform placement should be jointly designed rather than treated in isolation.}
Beyond FC and PC arrays, two further hardware families have become increasingly relevant for UAV payloads.
Holographic and RHS architectures replace the conventional phased array with a thin, low-profile metasurface whose element weights can be tuned in both amplitude and phase, offering a favorable trade-off between aperture size, weight, and beamforming flexibility that suits the planar form factor of a UAV payload.
Beamspace or lens-based architectures exploit the angular sparsity of the mmWave air-to-ground channel to map a large antenna aperture onto a small number of RF chains through a discrete lens array~\cite{hoang2023gradient,hoang2025rootmusic,hoang2026jointdoa}, which reduces both the number of phase shifters and the calibration burden.
Fig.~\ref{fig:uav_architecture} shows the four architecture families that can be integrated into a UAV BS. 
The figure highlights how RF-chain and phase-shifter flexibility differs across the four options, motivating the architecture choices reviewed below.

{Among the foundational architecture studies, the authors in~\cite{du_energy-saving_2020} consider a UAV that acts as an aerial BS and serves multiple single-antenna ground users through a massive MIMO array. They derive a closed-form rate expression and show that, under an LoS-dominant channel, the transmit power required to meet the rate target of each user decreases inversely with the number of UAV antennas. The optimal UAV location that minimizes this power reduces, for a path-loss exponent of two, to the SINR-weighted average of the user positions. Unlike~\cite{du_energy-saving_2020}, which optimizes the hovering location of the UAV, the work in~\cite{zhou_energy_2019} focuses on the transceiver itself and formulates an energy-efficient WMMSE power allocation problem for a UAV equipped with a broadband polarized tapped-delay array, which is solved by iterative convex quadratic programming. Both works evaluate a given architecture. In contrast, the work in~\cite{wang_beamforming_2020} compares architectures for a swarm of UAV BSs by means of stochastic geometry and shows that HBF consumes less circuit power than analog-only beamforming (ABF) at a comparable capacity for realistic UAV densities. Because multi-beam HBF reduces the number of UAVs needed to cover a given area, it reduces the total hovering and circuit power by up to $4.5\times$ with five parallel beams. With two beams, it saves 60\% of the power at the cost of only a 6\% ergodic-capacity loss at the optimal altitude, and at a low altitude of 10\,m it even improves the capacity by up to 33\% over ABF. The above works assume ideal phase control. Different from them, the work in~\cite{zhang_uav-bs_2023} accounts for quantized phase shifters at a UAV BS and minimizes the total transmit power under a minimum-rate constraint by Riemannian gradient descent on the constant-modulus manifold. The saving of the hybrid architecture over the fully digital one is further confirmed in two mission-specific deployments, namely a UAV-assisted network for disaster relief~\cite{sharma_power_2024} and a UAV inspection communication system for power lines~\cite{minzheng_energy_2021}, where the saving persists even for moderate array sizes.}
{The above designs mainly target the transmit power or the energy efficiency. Different from them, the work in~\cite{van2026conflict} targets the computational cost of the beam-selection stage in a multi-user mmWave UAV BS. The authors adopt greedy sector beam selection as the low-complexity default and activate a candidate-refinement stage together with regularized ZF precoding only for the roughly $17\%$ of channel realizations that beam-training indicators flag as conflict-prone. This selective refinement recovers most of the gain of always-on refinement at a complexity close to that of the greedy scheme.}

{Beyond conventional phased arrays, the authors in~\cite{chen_hybrid_2021,chen_joint_2022} adopt discrete lens arrays (DLAs) for beamspace precoding at a UAV BS and jointly optimize the UAV altitude, the beam selection, and the digital precoder by PDD. The beam-domain mapping reduces the number of required RF chains at the cost of a lens aperture that must fit within the payload budget. Instead of a lens, the work in~\cite{sheemar_joint_2026} replaces the phased-array payload of the UAV with an RHS and alternates a ZF digital stage with gradient-ascent updates of the holographic weights and of the UAV location, which improves the sum rate monotonically. The gain over the benchmark schemes originates from the dense and low-profile tunable aperture that suits a UAV payload. The work in~\cite{geng_mitigating_2025} also employs an RHS, but it focuses on the processing cost. The authors partition the surface into sub-blocks and exploit intra-block constructive interference in order to reduce the complexity of symbol-level precoding while retaining most of the array gain.}

Three observations follow from the above works.
First, UAV altitude directly controls the angular spread seen by the ground-user cluster, coupling platform placement to channel geometry in a way that has no direct counterpart in fixed terrestrial or orbital systems. 
Second, beamspace and lens-based architectures are attractive at mmWave frequencies precisely because the angular sparsity of air-to-ground links permits aggressive RF-chain reduction.
Third, holographic and reconfigurable surface designs add a further degree of aperture reuse well suited to the thin, planar form factor of UAV payloads, provided that the surface weight and the DC power budget are respected.

\subsection{\textcolor{black}{Mobility-Aware Beam Tracking and Management}}
\label{ssec:c2}

\begin{figure}[!t]
%
\centering
\includegraphics[width=\linewidth]{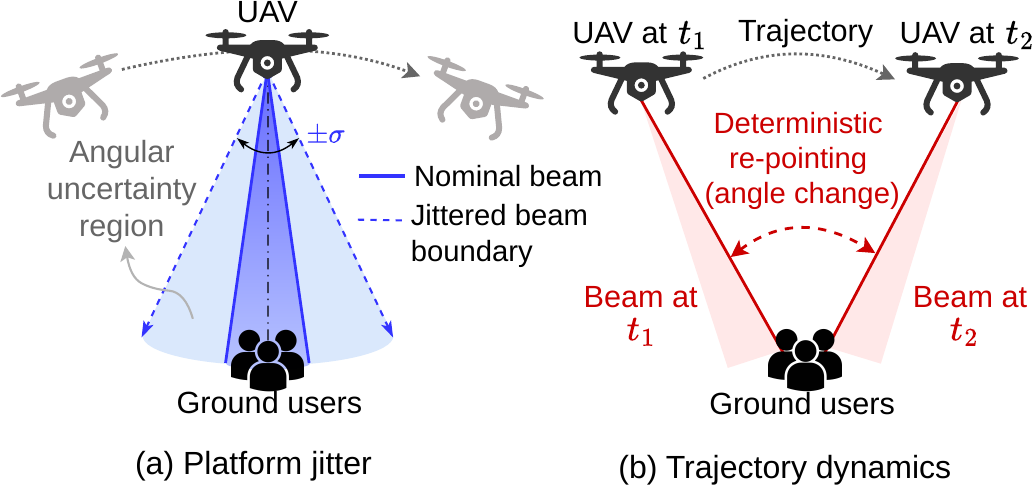}
\caption{Two mobility-induced impairments in UAV HBF. 
(a) Platform jitter produces a fast, stochastic angular spread around the nominal beam direction, motivating wide-beam or sensing-driven robust designs.
(b) Trajectory dynamics produce a slower, deterministic shift of the nominal pointing angle, motivating beam tracking or joint
trajectory-beamforming optimization.}
\label{fig:uav_jitter}
\vspace{-0.25cm}
\end{figure}

UAV mobility introduces two impairments that differ qualitatively from one another and from the Doppler and CSI-aging effects encountered in LEO HBF.
Fig.~\ref{fig:uav_jitter} illustrates the two impairments side by side. 
Platform jitter is a high-frequency, near-zero-mean stochastic process driven by motor vibration, propeller imbalance, and wind gusts, and it produces instantaneous angular errors of a few degrees that change faster than a typical channel-estimation or beam-update cycle. 
On the other hand, trajectory dynamics are lower-frequency and largely deterministic angular variations caused by waypoint following or coverage-area changes, and they set the timescale on which the analog beam must be refreshed rather than the timescale on which it must be made robust. 
Because narrow analog beams maximize array gain but are intolerant of pointing error, the common engineering response to jitter is to deliberately widen the beam, predict the channel ahead of time using motion or sensing information, or both. 
For trajectory dynamics, the natural response is instead to track the beam through machine-learning-assisted prediction or to jointly optimize the flight path together with the beamformer so that the beam-update rate itself becomes part of the design objective.

{The authors in~\cite{chen_adaptive_2024} consider a UAV that serves ground users through a mmWave array subject to mechanical jitter. They show that a circularly symmetric wide beam is suboptimal because the vertical and the horizontal jitter distributions differ, and they jointly optimize the azimuth and the elevation beam ranges in order to maximize the average rate under asymmetric jitter. Following a similar wide-beam principle, the works in~\cite{liu_deployment_2023,liu_robust_2022-1} design a chirp-sequence analog beam that covers the jitter-induced angular interval of each user, followed by ZF precoding over the jitter-averaged channel. The design maximizes the minimum user rate and is validated on a planar-array UAV testbed, where beam broadening outperforms narrow-beam refinement once the jitter exceeds approximately one degree. Departing from beam broadening, the work in~\cite{tang2026ssb} replaces explicit CSI feedback by a sensing-assisted predictive framework based on the synchronization signal block (SSB). A hierarchical pipeline of two-dimensional range-velocity profiling and beamspace multiple signal classification (MUSIC) estimates the state of the UAV from the periodic and always-on SSB. An extended Kalman filter (EKF) then tracks this state between sparse bursts with a covariance correction for maneuver uncertainty, and the predicted correlation matrices drive a robust multi-user HBF that is solved by SCA and alternating minimization. As standardized sensing signaling matures, such designs can reduce the pilot overhead required for jitter and mobility robustness.}

{For tracking against trajectory dynamics, the authors in~\cite{chiang_machine-learning_2021} consider mmWave links among multiple UAVs and apply tabular Q-learning. They model the beam transitions as a Markov decision process (MDP) and pre-load candidate beams in order to reduce the sweep overhead while maximizing the SINR at the digital stage. The work in~\cite{sim_unscented_2023} addresses the same tracking problem, but it replaces learning by an unscented Kalman filter (UKF) tracker embedded in an HBF framework for 5G NR MIMO. Although this framework is not dedicated to an aerial platform, it applies directly to a UAV link whose nominal pointing angle drifts slowly, and it enables low-latency multi-user and multi-stream tracking. Rather than filtering radio measurements only, the work in~\cite{sugimoto_hybrid_2025} exploits onboard sensors on UAV-to-UAV links and switches between LiDAR-camera fusion at short range and global navigation satellite system (GNSS) steering at long range in order to keep the swarm links stable. On a 60\,GHz $16\times16$ array, fusion-based steering achieves 11.98\,bit/s/Hz, whereas GNSS-only steering and codebook sweeping achieve 9.54 and 10.52\,bit/s/Hz at short range, respectively. Switching at a 19\,m threshold adds a further 8.5\% over fusion alone once the range extends to 30\,m. Coarser side information is already sufficient in~\cite{wu_location_2020}, where the beam codewords of an aerial user are selected from the global positioning system (GPS) coordinates of the UAV and of the user before ZF precoding. Instead of the position, the authors in~\cite{ren_machine_2019} learn a stochastic beam-selection policy for a UAV mmWave link by cross-entropy optimization, whereas the work in~\cite{hong_deep_nodate} predicts the next analog precoder of a moving UAV from two successive pilot frames through a deep double-pilot scheme.}

{Several works instead treat the trajectory of the UAV as an optimization variable. The work in~\cite{zhou_joint_2021} jointly designs the flight trajectory and the HBF of a UAV BS with FC and sub-connected arrays in order to maximize the weighted sum rate. It combines a Lagrangian dual transform with an alternating direction method of multipliers (ADMM)-based alternation between an SCA trajectory subproblem and a manifold-gradient beamforming subproblem. While~\cite{zhou_joint_2021} targets the sum rate, the authors in~\cite{silvirianti_sub-connected_2023} target the energy efficiency and pair sub-connected HBF with DRL-based trajectory optimization under battery and QoS constraints, where the hardware saving of the sub-connected architecture is used to extend the flight time.}


Across these designs, the trajectory-HBF coupling resolves into two timescales.
The analog beam must be refreshed within the beam-coherence time, a quantity set by the link geometry rather than by the array alone: it grows with beamwidth and range but shrinks with transverse speed~\cite{va_impact_2017}. 
{The digital precoder and the trajectory follow a slower scheduling timescale, which mirrors the analog-digital separation measured on LEO payloads in Section~\ref{sec:leo}. Conflating the two timescales either over-broadens the analog beam, which sacrifices array gain, or refreshes it too rarely, which incurs misalignment.}
\subsection{\textcolor{black}{Network-Level Design: Multi-UAV Cooperation and Relaying}}
\label{ssec:c3}

\begin{figure*}[!t]
\centering
\begin{subfigure}{.485\linewidth}
  \centering
  \includegraphics[width=\linewidth]{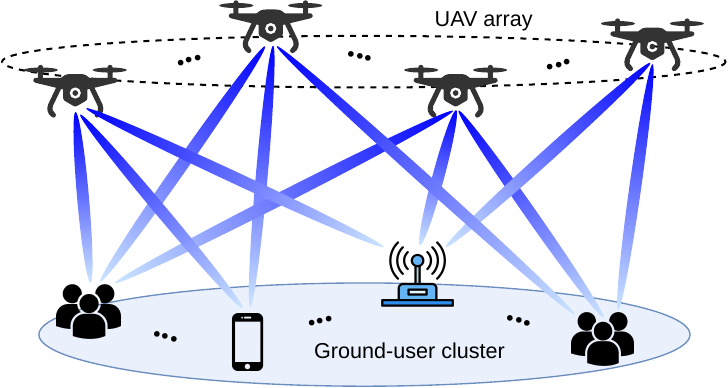}  
  \subcaption{Multi-UAV distributed virtual array}
\end{subfigure}
\begin{subfigure}{.465\linewidth}
  \centering
  \includegraphics[width=\linewidth]{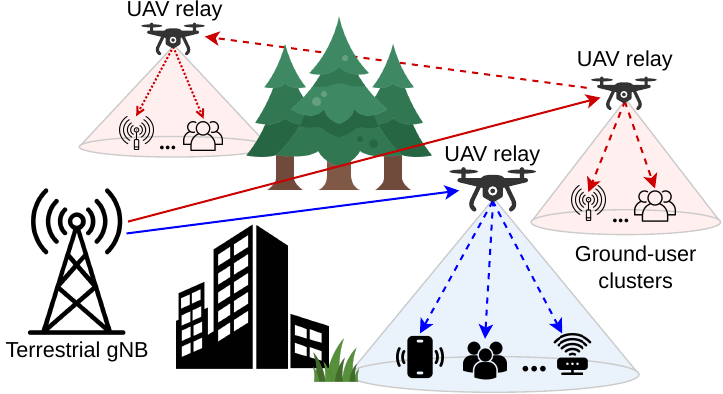}  
  \subcaption{UAV relays between a base station and blocked users}
\end{subfigure}
\caption{Cooperative UAV HBF topologies. 
(a) Multiple UAVs form a distributed virtual array, jointly beamforming toward a ground-user cluster. 
(b) UAVs relay traffic between a fixed base station and
users that lack a direct LoS path, requiring HBF design across multiple cascaded wireless hops.}
\label{fig:uav_multi_relay}
\end{figure*}

A single UAV BS is limited by its own antenna aperture and battery budget, which motivates two complementary cooperative strategies.
In the first, several UAVs jointly form a distributed virtual antenna array, in which each UAV contributes a subset of antennas to a combined beamformer that can offer array gain and interference management beyond the reach of any individual platform. In the second, UAVs act as aerial relays that forward traffic between a fixed BS and ground users who would otherwise suffer from blockage or excessive path loss. Both strategies introduce design issues that are largely absent from single-UAV HBF.
Fig.~\ref{fig:uav_multi_relay} shows the two cooperative topologies.
Distributed cooperation requires maintaining phase coherence across UAVs that are physically separated and driven by independent local oscillators, while relay-assisted HBF requires designing two cascaded wireless hops, each with its own analog and digital stage, and often requires deciding where to position the relay rather than only how to beamform from a fixed location.
In multi-UAV cooperation scenarios, the authors in~\cite{zhu_multi-uav_2022} formulate a joint UAV positioning, user clustering, and HBF problem for multiple mmWave UAV BSs. Under an ideal-beam-pattern approximation, the deployment and the clustering decouple from the beamforming, which yields a two-stage solution based on SCA and user association whose performance is close to that of fully digital beamforming. Whereas~\cite{zhu_multi-uav_2022} treats the UAVs as separate transmitters, the work in~\cite{wang_cooperative_2023} models the distributed UAVs as a single virtual transmitter and proposes a residual error minimization (REM) scheme that alternates LoS-exploiting analog alignment with digital updates that suppress inter-user interference across the virtual array. The work in~\cite{dong_joint_2025} extends multi-UAV cooperation to a secure ISAC setting and jointly optimizes the HBF, the artificial-noise (AN) injection, and the trajectories. The digital precoders and the noise covariance are learned by proximal policy optimization (PPO), and the fully digital solution is then decomposed into hybrid form by low-complexity matrix factorization, which separates long-timescale learning from short-timescale precoder decomposition. For an $8\times8$ array serving four legitimate UAVs against three eavesdroppers, PPO improves the sum secrecy rate by 39.3\% over the strongest benchmark learner.

In relay-assisted scenarios, the authors in~\cite{mir_relay_2022} propose a two-stage wideband design for a UAV relay, in which the analog precoder is obtained from the frequency-averaged angular support while the per-subcarrier digital precoders correct the residual gain variation. Whereas~\cite{mir_relay_2022} keeps the relay position fixed, the work in~\cite{koc_efficient_2023} also optimizes it and designs the HBF and the placement of a UAV relay in a dual-hop massive MIMO IoT network. The RF stage is obtained from slow angular information, the baseband stage by SVD, and the placement by particle swarm optimization (PSO). The work in~\cite{mahmood_pso-based_2022} extends~\cite{koc_efficient_2023} to the multi-user case through a joint hybrid precoding and positioning (JHPP) scheme that combines RZF digital precoding with PSO placement, and it reports a higher aggregate throughput than relays at fixed positions. Instead of the relay position, the authors in~\cite{belaoura_performance_2022} optimize the user association in a multi-BS system in which the UAV relay bridges urban LoS gaps, so that the sum rate is maximized while blockage outages are avoided. The above works assume accurate CSI. In contrast, the work in~\cite{wang_coordinated_2021} predicts the outdated effective channel of the UAV relay link at a coordinated BS by Gaussian process machine learning (GPML) and compensates it by a maximum-SINR scheme in order to curb inter-BS interference.

\subsection{\textcolor{black}{Sensing Capability and Reconfigurable Surfaces}}
\label{ssec:c4}

\begin{figure*}[!t]
\centering
\begin{subfigure}{.45\linewidth}
  \centering
  \includegraphics[width=.95\linewidth]{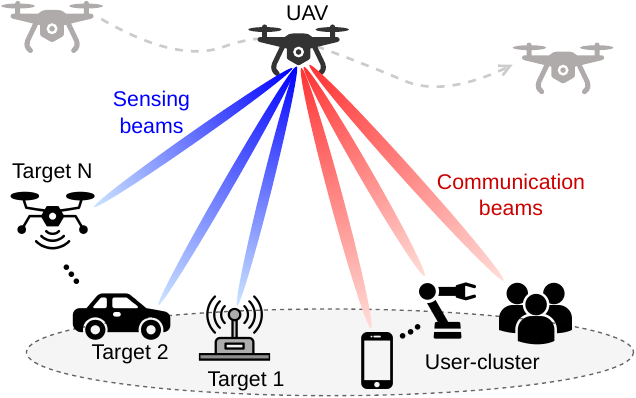}  
  \subcaption{UAV ISAC: shared communication and sensing beams}
\end{subfigure}
\begin{subfigure}{.45\linewidth}
  \centering
  \includegraphics[width=.95\linewidth]{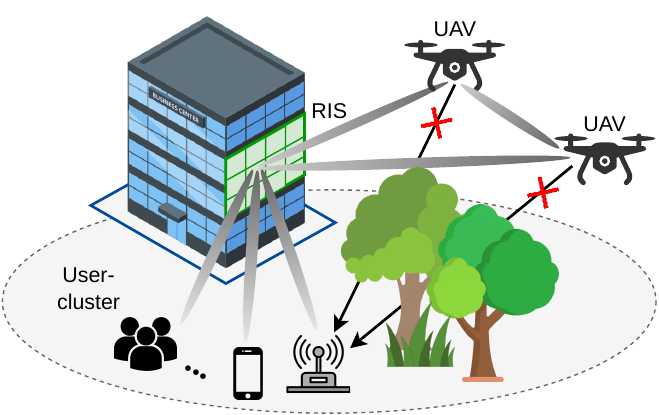}  
  \subcaption{RIS-assisted UAV coverage restoration}
\end{subfigure}
\caption{Sensing and reconfigurable-surface extensions of UAV HBF.
(a) A UAV array splits its beamforming resources between communication beams toward ground users and sensing beams toward sensing targets, exploiting its own mobility as a diversity dimension. 
(b) UAVs cooperate with a RIS to restore coverage to ground users blocked from the direct LoS path.}
\label{fig:uav_isac_ris}
\vspace{-0.25cm}
\end{figure*}

The combination of an elevated vantage point, a largely LoS-dominant channel, and a controllable trajectory makes a UAV a natural platform for ISAC. A UAV-mounted array can dedicate part of its beamforming resources to communication and part to radar-like sensing, and the sparsity of the air-to-ground channel means that the angular information extracted from sensing is often accurate enough to replace, or substantially reduce, conventional CSI feedback. A second and complementary direction equips the UAV, or a structure observed by the UAV, with an RIS that passively or actively reflects the signal toward users that are otherwise blocked. Because the RIS phase profile, the hybrid precoder, and the UAV trajectory or position jointly determine the end-to-end channel, RIS-assisted UAV HBF is typically a higher-dimensional design problem than either UAV-only or ground-RIS designs.
Fig.~\ref{fig:uav_isac_ris} depicts the two complementary functions discussed in this subsection: a UAV performing joint communication and sensing toward ground users and sensing targets, and UAVs cooperating with a RIS to extend coverage around an obstacle.

{In airborne ISAC, the authors in~\cite{salih_spectral_2025} propose a low-complexity model for a UAV that serves multiple ground users and senses simultaneously, together with a scattering-aware path-loss model. They show that, at the altitude that balances the LoS probability against the path loss, the scheme attains a high spectral efficiency at an acceptable sensing-outage probability. While~\cite{salih_spectral_2025} places the ISAC functionality on the UAV itself, the work in~\cite{wu_joint_2025} uses the UAV as an aerial platform that carries an active RIS and jointly optimizes the hybrid precoders of the terrestrial BS and the active-RIS beamforming matrix in a dual-function radar-communication (DFRC) system by alternating WMMSE-based PDD and SDR. The design maximizes the user SINR under a radar-beampattern constraint, and the amplification of the active RIS improves the sensing-communication trade-off with respect to a passive RIS. Sensing is exploited differently in~\cite{zhang_sensing-assisted_2024}, where the AoA estimated at the ground BS replaces the uplink CSI feedback of the UAV. The receive HBF is obtained by SDR and manifold optimization, and the UAV trajectory is jointly refined by SCA, because the LoS dominance makes the UAV AoA a near-sufficient statistic for the analog design. The authors in~\cite{xu_location-based_2022} instead quantify the sensing accuracy explicitly. They derive the Cram\'er-Rao lower bound (CRLB) for the joint radar estimation in a system in which a UAV cooperates with an unmanned ground vehicle (UGV), and they balance the rate against the sensing accuracy by difference-of-convex and QoS-aware power allocation.}
{At the array level rather than at the precoder level, the work in~\cite{jiang20263d} proposes a spherical directly-connected antenna array (DCAA) for low-altitude UAV-swarm ISAC, which eliminates the phase shifters by connecting the elements of planar subarrays distributed over a spherical surface directly. Relative to a conventional HBF array, the DCAA lowers the hardware cost and provides a more uniform three-dimensional angular resolution, which improves both the sensing resolution and the spectral efficiency when the targets drift away from the array boresight.}

{For coverage extension, the authors in~\cite{kakati_hybrid_2025} deploy simultaneously-transmitting-and-reflecting RIS (STAR-RIS) in order to extend the mmWave coverage of a UAV across indoor-outdoor environments. The precoder, the phase profiles, and the trajectory are jointly optimized under a secrecy constraint by a generative-AI and differential-privacy enhanced soft actor-critic (GAIDP-SAC) agent, whose synthetic channels improve the sample efficiency near the indoor-outdoor boundary. Similarly to~\cite{kakati_hybrid_2025}, the work in~\cite{sun_energy-efficient_2022} also couples the RIS with the UAV, but it considers a multi-layer terrestrial-aerial network and cooperatively optimizes the hybrid precoder of the BS, the passive-RIS phase-shift matrix, and the trajectory of the UAV relay under a minimum-secrecy-rate constraint. The SDR precoder updates, the gradient-projection RIS updates, and the SCA trajectory refinement alternate in order to maximize the secure energy efficiency.}

Notably, joint UAV trajectory and RIS phase-shift optimization creates a three-way coupled problem involving the precoder, the RIS phase, and the UAV position, markedly more complex than either ground-RIS or UAV-only design.
{Existing solutions handle this coupling through alternating optimization with DRL or SCA sub-solvers, and the convergence is generally only local. Geometric initialization heuristics, such as placing the UAV on the line that connects the BS and the served user cluster, are observed to improve the solution quality in practice.}

\subsection{Security and Multiple Access}
\label{ssec:c5}

\begin{figure*}[!t]
\centering
\begin{subfigure}{.32\linewidth}
  \centering
  \includegraphics[width=\linewidth]{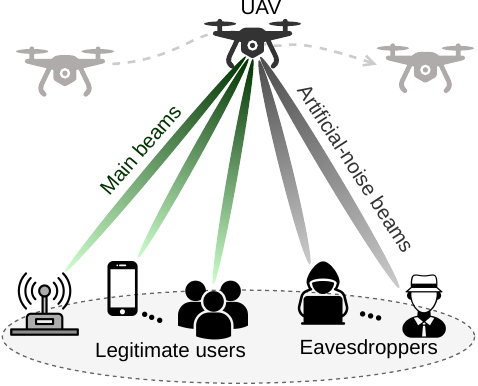}  
  \subcaption{UAV-aided PLS}
\end{subfigure}
\begin{subfigure}{.66\linewidth}
  \centering
  \includegraphics[width=\linewidth]{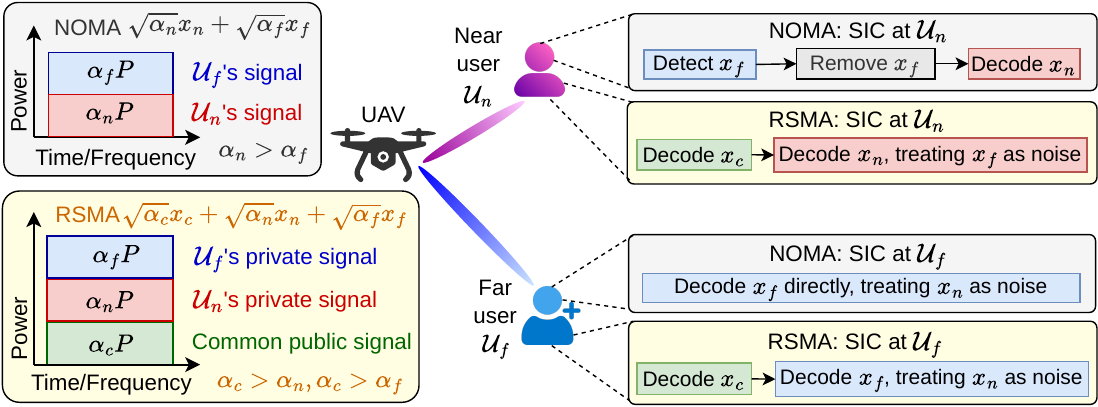}  
  \subcaption{NOMA/RSMA-based UAV}
\end{subfigure}
\caption{Security and multiple-access mechanisms layered on UAV HBF.
(a) The analog stage steers the main beam toward legitimate users while the digital stage injects AN toward suspected eavesdroppers. 
(b) A single analog beam can serve several users simultaneously through power-domain NOMA or through RSMA, which splits each
message into a common and a private stream.}
\label{fig:uav_security_noma}
\vspace{-0.3cm}
\end{figure*}
The elevated position of a UAV BS creates a near-broadcast downlink geometry, since ground eavesdroppers at almost any location receive a non-trivial signal from the high-altitude transmitter. 
HBF is well suited to PLS because the analog stage can steer the main beam toward legitimate users while the digital stage injects artificial noise into the null space of the legitimate channel, and because the UAV's own mobility provides a further, geometric
degree of freedom: the platform can simply move away from suspected eavesdropper locations. A separate but related design question concerns how a single hybrid beam serves multiple users efficiently. NOMA groups users by channel correlation and separates
them in the power domain using successive interference cancellation (SIC), exploiting the strong LoS gain differences typical of air-to-ground links, while RSMA splits each user's message into a common part decodable by all users and a private part decodable only by its intended recipient, offering a more graceful degradation when the channel ordering among users is uncertain.
Fig.~\ref{fig:uav_security_noma} summarizes the two themes of this
subsection. Fig.~\ref{fig:uav_security_noma}(a) shows a UAV steering its main beams toward
legitimate users while injecting artificial-noise beams toward suspected
eavesdroppers, whereas 
Fig.~\ref{fig:uav_security_noma}(b) shows the NOMA and RSMA multiple-access structures sharing the same analog beam.

{On PLS, the authors in~\cite{dong_hybrid_2023} maximize the secrecy rate of a massive MIMO UAV network. With perfect eavesdropper CSI, they obtain the optimal transmit covariance by generalized eigenvalue decomposition (GED) and match it into hybrid form by OMP, whereas for unknown eavesdropper CSI a network trained on legitimate-user CSI maps directly to a near-optimal HBF matrix. An earlier work of the same group~\cite{dong_hybrid_2021} alternately optimizes the digital and the analog precoders by SCA under a per-antenna power constraint. Rather than relying on eavesdropper CSI, the work in~\cite{niu_physical_2023} combines radar-assisted beam selection that steers energy away from the estimated eavesdropper positions with difference-of-convex power allocation and with a path-planning step that routes the UAV so that the eavesdroppers are not illuminated. The mobility of the UAV is exploited as a secrecy mechanism in~\cite{dong_joint_2025} as well, where trajectory planning provides geometric secrecy in multi-UAV ISAC, and in~\cite{sun_energy-efficient_2022}, where the RIS phase shifts and the trajectory reshape the channel so that the eavesdropper SINR falls below a secrecy threshold.}

{On multiple access, the authors in~\cite{feng_hybrid_2021} combine power-domain NOMA with HBF in a UAV wireless-powered mobile-edge-computing (MEC) network. The three-dimensional placement of the UAV is optimized by polyhedral annexation, whereas the HBF and the offloading are optimized by deep deterministic policy gradient (DDPG), and the intra-cluster interference is absorbed by the NOMA SIC receiver rather than by the digital precoder. NOMA is also adopted in~\cite{pierucci_hybrid_2022} for the uplink of a UAV relay in 6G networks, where MUSIC direction-of-arrival estimation drives the hybrid precoder and the relay-side SIC handles the residual interference. Instead of a single cluster, the work in~\cite{hevesli_hybrid_2025} groups the users of multiple UAVs by angular proximity into a clustered NOMA (C-NOMA) scheme, serves each cluster by one analog beam, and solves the joint HBF and intra-cluster power allocation as an MDP by multi-agent PPO, which improves the energy efficiency by up to 37.7\% over learning-based baselines without HBF. Because the performance of NOMA depends on the channel ordering, the work in~\cite{wang_deep_2025} turns to RSMA and applies a deep residual network to hybrid RSMA precoding for a UAV link under jamming, where every output is projected onto the feasible power-rate set so that a separate feasibility step is removed. The authors in~\cite{everett_energy_2024} analyze a one-layer RSMA hybrid precoder for mmWave 6G. Since space-division multiple access (SDMA) is the special case with no common stream, the design is at least as energy-efficient as an SDMA precoder by construction, and its advantage grows as the user channels become neither fully aligned nor fully orthogonal.}

NOMA-based UAV HBF exploits the strong LoS gain difference between near and far users to enable SIC, but it requires the digital precoder to maintain carefully controlled beam-power ratios that jitter or trajectory drift can disrupt. RSMA's partial interference management is more forgiving of channel-ordering uncertainty, at the cost of an additional common-stream precoder that must remain decodable by all served users simultaneously, which is a distinct constraint on the analog beam that complicates joint design within a reduced-RF-chain architecture.

\subsection{Summary and Lessons Learned}
\label{ssec:uav_summary}

\begin{table*}[!t]
\centering
\caption{\normalfont\scshape Summary of Representative HBF Architectures and Design Techniques for UAV Systems}
\label{tab:uav_survey_comparison}
\renewcommand{\arraystretch}{1.15}
\setlength{\tabcolsep}{2.6pt}
\scriptsize
\begin{tabular}{|c|c|l|l|l|l|}
\hline
\cellcolor{myblue}{\textbf{Reference}} &
\cellcolor{myblue}{\textbf{Architecture}} &
\multicolumn{1}{c|}{\cellcolor{myblue}{\textbf{Methodology}}} &
\multicolumn{1}{c|}{\cellcolor{myblue}{\textbf{Platform Coupling}}} &
\multicolumn{1}{c|}{\cellcolor{myblue}{\textbf{Key Enabling Technique}}} &
\multicolumn{1}{c|}{\cellcolor{myblue}{\textbf{Objective}}} \\
\hline\hline
\multicolumn{6}{|l|}{\cellcolor{gray!15}\textit{System Architecture and Precoding Design}} \\
\hline
\cite{du_energy-saving_2020} & FC & Closed-form & Altitude optimization & Closed-form rate analysis & SR \\
\hline
\cite{zhou_energy_2019} & FC & WMMSE & Fixed & Tapped-delay polarized array & EE \\
\hline
\cite{wang_beamforming_2020} & FC/ABF & Stochastic geometry & Density optimization & ABF versus HBF capacity trade-off & SR / EE \\
\hline
\cite{zhang_uav-bs_2023} & FC & Manifold optimization & Fixed & Riemannian gradient, quantized phase shifters & Power minimization \\
\hline
\cite{sharma_power_2024} & PC & AO & Fixed & Disaster-relief energy-constrained HBF & EE \\
\hline
\cite{minzheng_energy_2021} & PC & AO & Fixed & Power-line inspection HBF & EE \\
\hline
\cite{van2026conflict} & FC & Greedy + RZF & Fixed & Conflict-triggered beam refinement & SE \\
\hline
\cite{chen_hybrid_2021,chen_joint_2022} & Beamspace & PDD & Altitude optimization & DLA beamspace & SR \\
\hline
\cite{sheemar_joint_2026} & Holographic & AO + gradient ascent & 3D position optimization & RHS holographic beamforming & SR \\
\hline
\cite{geng_mitigating_2025} & Holographic & Sub-block & Fixed & Sub-block holographic precoding & SR / EE \\
\hline\hline
\multicolumn{6}{|l|}{\cellcolor{gray!15}\textit{Mobility-Aware Beam Tracking and Management}} \\
\hline
\cite{chen_adaptive_2024} & FC & Beam-range optimization & Jitter model & Asymmetric jitter-robust HBF & Average SR \\
\hline
\cite{liu_deployment_2023,liu_robust_2022-1} & FC & Max-min & Jitter + deployment & Chirp-sequence wide beam + ZF & MMR \\
\hline
\cite{tang2026ssb} & FC & SCA + alternating minimization & Mobility/EKF & SSB sensing + predictive robust HBF & SR (robust) \\
\hline
\cite{chiang_machine-learning_2021} & FC & Q-learning & Mobility & MDP beam tracking & SINR \\
\hline
\cite{sim_unscented_2023} & FC & UKF & Trajectory & Unscented Kalman beam tracking & Multi-stream SR \\
\hline
\cite{sugimoto_hybrid_2025} & FC & Sensor fusion & UAV-UAV & LiDAR-GNSS fusion steering & Link stability \\
\hline
\cite{wu_location_2020} & FC & Location-aided & Fixed & GPS-assisted beam selection & SR \\
\hline
\cite{ren_machine_2019} & FC & Cross-entropy ML & Fixed & Stochastic beam-index learning & SR \\
\hline
\cite{hong_deep_nodate} & FC & DNN & Velocity & Double-pilot channel prediction & SR \\
\hline
\cite{zhou_joint_2021} & FC/PC & AO + ADMM & Trajectory optimization & Lagrangian dual + ADMM & SR \\
\hline
\cite{silvirianti_sub-connected_2023} & PC & DRL & Trajectory optimization & Sub-connected + DRL trajectory & EE \\
\hline\hline
\multicolumn{6}{|l|}{\cellcolor{gray!15}\textit{Network-Level Design: Multi-UAV Cooperation and Relaying}} \\
\hline
\cite{zhu_multi-uav_2022} & FC & AO + SCA & Multi-UAV deployment & Positioning, clustering, and beamforming decoupling & SR \\
\hline
\cite{wang_cooperative_2023} & FC & REM & Multi-UAV cooperation & Residual error minimization & SR \\
\hline
\cite{dong_joint_2025} & FC & PPO + matrix factorization & Multi-UAV trajectory & DRL + AN + ISAC security & Secrecy \\
\hline
\cite{mir_relay_2022} & FC & Two-stage AO & Relay position & Wideband analog + per-subcarrier digital & SR \\
\hline
\cite{koc_efficient_2023} & FC & SVD + PSO & Relay altitude & SVD baseband + PSO placement & SR \\
\hline
\cite{mahmood_pso-based_2022} & FC & RZF + PSO & Relay position & JHPP, multi-user RZF & SR \\
\hline
\cite{belaoura_performance_2022} & FC & User association & Relay hover & LoS blockage-aware association & SR \\
\hline
\cite{wang_coordinated_2021} & FC & GPML & Relay motion & Gaussian process CSI prediction & SINR \\
\hline\hline
\multicolumn{6}{|l|}{\cellcolor{gray!15}\textit{Sensing Capability and Reconfigurable Surfaces}} \\
\hline
\cite{salih_spectral_2025} & FC & AO & Altitude optimization & Scattering path-loss ISAC model & SE / outage \\
\hline
\cite{wu_joint_2025} & FC & WMMSE-PDD, SDR & Fixed & Active RIS-assisted DFRC & SINR (communication / radar) \\
\hline
\cite{zhang_sensing-assisted_2024} & FC & SDR, manifold optimization, SCA & Trajectory optimization & AoA sensing replaces CSI feedback & SR \\
\hline
\cite{xu_location-based_2022} & FC & Difference-of-convex programming & UAV-UGV cooperation & Radar-aided location HBF + CRLB & QoS power \\
\hline
\cite{jiang20263d} & DCAA & Array design & UAV swarm & Spherical phase-shifterless 3D array & SE / resolution \\
\hline
\cite{kakati_hybrid_2025} & FC & GAIDP-SAC & Trajectory optimization & STAR-RIS + genAI DRL + privacy & Secrecy \\
\hline
\cite{sun_energy-efficient_2022} & FC & AO (SDR+SCA) & Trajectory optimization & Multilayer RIS-UAV-BS PLS & EE with secrecy \\
\hline\hline
\multicolumn{6}{|l|}{\cellcolor{gray!15}\textit{Security and Multiple Access}} \\
\hline
\cite{dong_hybrid_2023} & FC & GED + DNN & Fixed & GED + DNN for unknown eavesdropper CSI & Secrecy \\
\hline
\cite{dong_hybrid_2021} & FC & AO + SCA & Fixed & SCA-based AN precoder & Secrecy \\
\hline
\cite{niu_physical_2023} & FC & Difference-of-convex programming & Path planning & Radar beam selection + path planning & Secrecy \\
\hline
\cite{feng_hybrid_2021} & FC & DDPG + polyhedral & 3D deployment & NOMA + MEC + DDPG & Computation rate \\
\hline
\cite{pierucci_hybrid_2022} & FC & MUSIC direction-of-arrival & Relay position & MUSIC direction-of-arrival + NOMA SIC relay & SR \\
\hline
\cite{hevesli_hybrid_2025} & FC & Multi-agent PPO & Multi-UAV & C-NOMA + multi-agent PPO & EE \\
\hline
\cite{wang_deep_2025} & FC & Deep residual network & Fixed & RSMA + constraint module & SR (jamming) \\
\hline
\cite{everett_energy_2024} & FC & Analytical & Fixed & One-layer RSMA EE analysis & EE \\
\hline
\end{tabular}
\end{table*}

Table~\ref{tab:uav_survey_comparison} consolidates the surveyed works by architecture, methodology, platform coupling, key enabling technique, and design objective. Across the five themes a consistent picture emerges, from which we distill the main lessons.

System architecture and precoding design (Section~\ref{ssec:c1}) remain dominated by FC and PC structures, with beamspace and lens designs gaining traction at mmWave frequencies and holographic or RHS payloads emerging as a weight- and power-efficient alternative. What distinguishes UAV HBF here is that platform placement is itself a first-class beamforming variable: in every category surveyed, UAV position or trajectory is either the primary optimization variable or a binding constraint on the feasible beamformer set, so frameworks that treat placement and beamforming as separate, sequential problems consistently underperform joint co-design, unlike terrestrial HBF with fixed nodes or LEO HBF with predictable orbital mechanics.

Mobility-aware beam tracking and management (Section~\ref{ssec:c2}) are handled by beam broadening or, more recently, sensing-driven prediction against high-frequency jitter, and by learning-based tracking or joint trajectory-beamforming co-optimization against slower trajectory dynamics. The recurring lesson is that a timescale hierarchy must be respected: jitter acts at well under ten milliseconds, trajectory updates from roughly a hundred milliseconds to seconds, and mission planning over minutes. 
{The analog beam should track the trajectory timescale, the digital beam should absorb the residual fast variation, and the trajectory should be planned so as to minimize the rate of analog-beam changes. Designs that conflate these scales drift toward either over-broad beams or excessive update overhead.}

Cooperative and sensing-aided designs (Sections~\ref{ssec:c3} and~\ref{ssec:c4}) extract array gain from treating distributed UAVs as a virtual array, with PSO a recurring relay-positioning tool, and exploit UAV mobility as a free diversity dimension for ISAC. A central lesson is that channel sparsity turns sensing into a practical CSI surrogate: because the LoS-dominant air-to-ground support is fixed almost entirely by the AoA at the ground user and the angle of departure at the UAV, location- and sensing-aided designs replace much of the instantaneous CSI feedback at far lower overhead, an advantage likely to grow as standardized sensing signaling matures. RIS integration, by contrast, couples precoder, surface phase, and UAV position into a three-way problem solved only locally by alternating optimization with DRL or convex-approximation sub-solvers. More broadly, learning is most effective when grounded in physical structure: the strongest schemes embed priors such as the sparse mmWave model, UAV kinematics, the SIC decoding order, or periodic sensing signals, making model-driven DRL and unfolding-based supervised learning more data-efficient and more transferable across altitudes and bands than black-box policies.

Finally, security and multiple access (Section~\ref{ssec:c5}) show that HBF enables PLS through joint beam steering and artificial-noise injection, with trajectory control as a complementary geometric secrecy mechanism, while NOMA, clustered NOMA, and RSMA each interact favorably with the analog beam in distinct ways. The key lesson is that energy efficiency and PLS are coupled rather than independent: UAV HBF is bounded below by battery endurance and above by the secrecy vulnerability of an elevated, near-broadcast transmitter, and artificial-noise injection spends power that would otherwise extend flight time. Few surveyed works optimize energy efficiency and secrecy rate under a unified objective, marking this as a priority for future work.

\section{Existing Challenges, Open Issues, \\and Future Directions}
\label{Sect:OpenChallenges}

The platform-aware review in Sections~\ref{sec:leo} and~\ref{sec:uav} shows that HBF for LEO satellite and UAV systems now rests on a mature methodological foundation, yet most reported designs are still validated in simulation, on a single platform, and under idealized RF hardware. Building on this observation, the present section collects the open challenges that the survey's taxonomy exposes. Each challenge cuts across both platform families and couples the beamforming problem to constraints, namely finite-resolution hardware, wideband beam squint, sensing integration, cross-layer coordination, ultra-wideband THz operation, and payload-realistic RF networks, that existing single-platform studies address only in isolation.

At a high level, four themes recur across these challenges before we develop them in the dedicated subsections that follow, namely hardware-aware HBF implementation, integrated sensing, communication, and beamforming, 3D multi-layer NTN, and THz HBF for NTN.
 
\subsection{Hardware-Aware HBF Implementation}
 
\subsubsection{Low-Resolution and Quantized RF Hardware}
Despite the increasing attention paid to hardware-constrained HBF for NTN systems, existing studies remain largely limited to low-resolution phase-shifter networks, TRPS, and nonlinear power-amplifier-aware precoding designs, as discussed in Section \ref{ssec:b1} through the representative works of \cite{you_hybrid_2022,you_massive_2022,kong_hybrid_nodate}. Meanwhile, the tutorial discussion in Section \ref{ssec:sysmodel} highlighted that practical NTN payloads must operate under stringent size, weight, power, and thermal (SWaP-T) constraints, where payload power consumption, insertion loss, calibration complexity, hardware reliability, and payload lifetime become primary design considerations rather than secondary implementation issues. Nevertheless, most existing HBF algorithms still assume a fixed RF hardware architecture with predetermined phase-shifter resolutions and ideal analog front-ends, optimizing communication performance only after the hardware configuration has been selected.

Finite-resolution phase control converts analog beamformer design from a continuous problem into a discrete one. If $N_{\mathrm{PS}}$ independently controlled phase shifters each support $Q$ states, exhaustive search spans $Q^{N_{\mathrm{PS}}}$ configurations. Existing methods either optimize a continuous-phase beamformer and project it onto a finite codebook \cite{shang2024ris}, or directly optimize discrete hardware states through search or learning-assisted methods \cite{lukito2024integrated,bao2026intelligent}. The former is computationally convenient but can introduce projection mismatch, whereas the latter preserves hardware feasibility but becomes difficult to scale with array size. Recent approaches based on heuristic deep reinforcement learning and conditional generative learning with Gumbel--Softmax quantization show that learning can reduce the online burden of discrete phase optimization \cite{bao2026intelligent}. However, these works mainly configure RIS or STAR-RIS states under fixed element topologies and resolutions rather than jointly controlling an active HBF front end and its digital precoder.

This limitation motivates adaptive hardware-aware HBF, in which phase resolution, RF connectivity, active switch or phase-shifter states, and analog--digital precoders are jointly optimized under time-varying NTN geometry and payload constraints. Although the physical bit depth of a conventional phase shifter is fixed, the minimum effective resolution, mixed-resolution activation pattern, and number of active high-resolution components required to meet a target can vary with scan angle, traffic demand, and available power. Relevant directions include mixed-resolution phase-shifter networks, switch-assisted architectures, adaptive RF-chain activation, low-resolution data converters, and learning-assisted discrete beam synthesis. The central problem is to determine the minimum resolution--connectivity profile required to satisfy spectral-efficiency, beam-reliability, latency, and energy-consumption targets without unnecessary RF complexity.
 
\subsubsection{Wideband Beam Squint and TTD-Based Architectures}
Wideband HBF in NTNs suffers from beam squint because frequency-flat phase shifters fail to match the frequency-dependent array response across OFDM subcarriers. The resulting beam displacement and edge-subcarrier gain loss increase with fractional bandwidth, array aperture, and scan angle. True-time-delay (TTD) elements provide frequency-dependent phase control, while delay-phase precoding (DPP) reduces the TTD count through a hybrid TTD--phase-shifter (PS) network \cite{Dai2022DPP}. 

\begin{figure*}[!t]
\centering
\includegraphics[width=.95\linewidth]{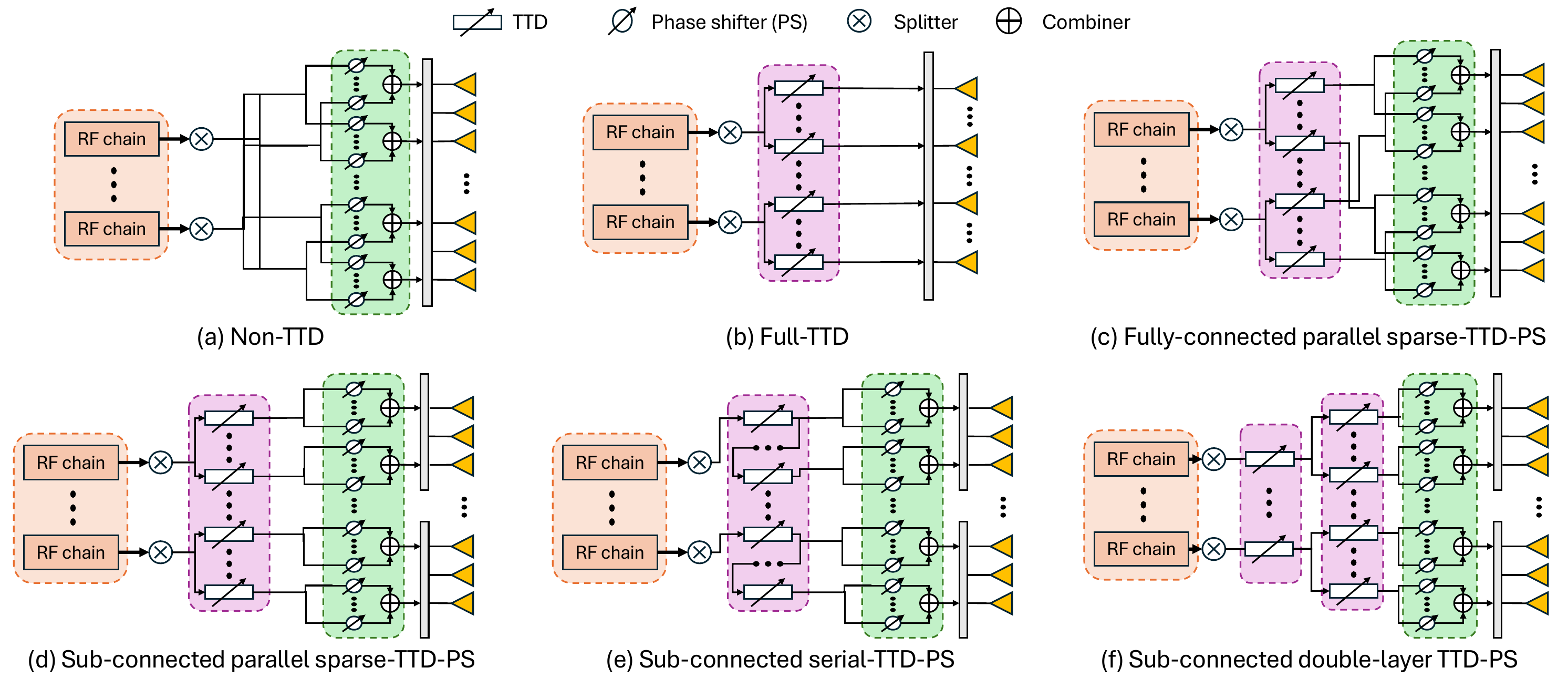}
\caption{Representative ULA-based HBF architectures for beam-squint mitigation in NTNs.}
\label{fig:TTD_Architectures_NTN}
\vspace{-0.25cm}
\end{figure*}

Fig.~\ref{fig:TTD_Architectures_NTN} summarizes representative beam-squint-mitigation architectures in terms of TTD density, RF connectivity, delay interconnection, and delay hierarchy. The non-TTD baseline in Fig.~\ref{fig:TTD_Architectures_NTN}(a) avoids delay hardware but suffers from beam squint, whereas Full-TTD in Fig.~\ref{fig:TTD_Architectures_NTN}(b) provides per-antenna delay control at a hardware and calibration cost that scales with the array size \cite{Monemi2026TTDComparison}. Parallel sparse-TTD--PS designs reduce the TTD count: the fully connected structure in Fig.~\ref{fig:TTD_Architectures_NTN}(c) preserves full-array access, while the sub-connected structure in Fig.~\ref{fig:TTD_Architectures_NTN}(d) reduces phase-shifter count, routing, and power at the expense of beamforming flexibility \cite{Dai2022DPP,Wang2025Beamfocusing}. Serial TTDs in Fig.~\ref{fig:TTD_Architectures_NTN}(e) reduce the required delay range per device through cumulative delays but introduce coupled control and insertion loss \cite{Wang2024TTDConfigurations}, whereas the double-layer structure in Fig.~\ref{fig:TTD_Architectures_NTN}(f) shares large-range delays among local small-range stages at the cost of hierarchical routing and calibration \cite{Sun2024DoubleLayer}. 
Since no architecture simultaneously minimizes squint, component count, delay range, loss, power, and control complexity, payload-level comparisons should incorporate hardware nonidealities and establish architecture crossover points under identical aperture, bandwidth, scan region, and radiated-power constraints \cite{Monemi2026TTDComparison,Nguyen2024JointDPP,Najjar2024HybridDPP,Schmid2013CalibrationCoupling}.

Beyond active TTD–PS networks, passive multi-beam arrays implemented by lenses \cite{Liu2018RotmanLens, park2022beam} or circuit-type beamforming networks \cite{guo2021circuitbfn} shift the RF complexity from per-element delay control to predefined beam ports, switching, and low-dimensional digital processing. These architectures can reduce the active analog-control burden, but they are not inherently frequency invariant, as experimentally observed in a recent lens-assisted HBF prototype~\cite{Wen2026GRINLensHBF}. For example, a lens/DFT beamspace response may migrate across beam ports over frequency. This migration can be compensated through frequency-aware multi-port selection or exploited for angular acquisition. The resulting open problem is the joint design of beam-port selection, switch connectivity, per-subcarrier digital precoding, and multi-timescale beam tracking under NTN Doppler, mobility, and SWaP-T constraints \cite{guo2021circuitbfn,wu2019lenssquint,wu2023greenmbaa}.


The selected architecture determines the feasible analog beamformer and its reconfiguration rate. Existing DPP and TTD-HBF methods approximate frequency-dependent fully digital precoders with a frequency-shared analog network \cite{Dai2022DPP,Nguyen2024JointDPP,Najjar2024HybridDPP,Yan2022DSFTTD}, while dynamic-subarray with fixed-TTD (DS-FTTD) and frequency-space mapping extend this principle to multiuser LEO links \cite{Lee2025LEOBeamSquint, Yan2022DSFTTD, Park2026RainbowBeamforming}. Their evaluation should move beyond isolated snapshots and ideal hardware to complete satellite passes or UAV trajectories, with quantized states, delayed CSI, residual Doppler, inter-beam interference, and different analog, digital, and scheduling update rates.

Beam squint also complicates channel acquisition because a center-frequency steering model cannot represent the spatial response across the full bandwidth. 
The authors in~\cite{Son2025TensorBeamSquint} develop constrained canonical polyadic (CP)- and Tucker-based tensor channel estimation for multi-satellite LEO links, whereas the work in~\cite{Lee2025LEOBeamSquint} employs statistical CSI for DS-FTTD beamforming and the work in~\cite{Park2026RainbowBeamforming} combines geometric and statistical CSI for rainbow-beam resource allocation.
However, channel estimation, wideband architecture selection, and resource allocation remain largely treated as separate problems. An open architecture-level problem is therefore the joint design of frequency-dependent angle–delay–Doppler acquisition, TTD or switch connectivity, and per-subcarrier digital precoding under quantized delay, insertion loss, and calibration constraints. The scheduling of these hardware states across different update timescales is addressed in Section \ref{Sec:AI_tracking}.
 
\subsubsection{Real-Time AI-Assisted Beam Tracking and Management} \label{Sec:AI_tracking}

Real-time AI-assisted beam management in NTN is a closed-loop, multi-timescale problem linking state acquisition, prediction, analog reconfiguration, digital precoding, and mobility management. UAV studies have developed individual components based on Kalman filtering, reinforcement learning, angular prediction, and sensor fusion \cite{sim_unscented_2023,chiang_machine-learning_2021,gao2022machine}, whereas NTN operation additionally involves delayed CSI, moving footprints, and constellation-scale coordination \cite{lin20215g,aristodemou2026multi,zhu2024beam,zhu2024beam_2}. These heterogeneous dynamics make fixed-period synchronous updates inefficient. Complementary work has investigated outage-aware hash-based opportunistic HBF for high-mobility mmWave UAV links \cite{le2026outageaware}, reflecting a parallel interest in reliability-aware beam adaptation under rapid mobility.
 
The need for differentiated update policies is particularly evident in HBF architectures. The analog RF beamformer tracks the dominant angular subspace under phase-shifter or subarray-switching constraints, whereas the digital precoder operates on the resulting low-dimensional effective channel to suppress residual multi-user interference and accommodate scheduling variations \cite{palacios_hybrid_2021, payami2018hybrid}. For Earth-fixed LEO coverage, the analysis in \cite{momani_beam_2025} indicates that analog beam realignment is typically required on a timescale of seconds, while zero-forcing digital precoding must be refreshed within tens of milliseconds to prevent appreciable sum-rate degradation. Moreover, lower orbital altitudes and larger antenna apertures require more frequent analog updates because their narrower ground footprints reduce the duration over which a selected beam remains valid \cite{momani_beam_2025}. Service-area-aware codebooks can partially alleviate this overhead by excluding steering vectors that do not illuminate the target coverage region \cite{palacios_hybrid_2021, momani_beam_2025}.
 
UAV-assisted links exhibit a different but similarly heterogeneous temporal structure. Mechanical jitter induces rapid stochastic angular perturbations, whereas trajectory evolution produces a slower and more predictable drift in the nominal pointing direction \cite{gao2022machine, chiang_machine-learning_2021, sim_unscented_2023}. A unified beam-management controller should therefore determine whether to retain the current configuration, update only the digital precoder, switch the analog beam, or initiate top-$K$ beam probing. This decision should jointly account for predicted array-gain degradation, CSI age, interference leakage, state uncertainty, pilot overhead, inference latency, and RF switching cost. Orbital ephemeris and localization information can provide low-overhead geometric predictions \cite{lin20215g, you_integrated_2024}, while pilot- and sensing-based observations are needed to correct errors caused by terminal mobility, blockage, platform jitter, and model mismatch. Although existing studies address individual components of this control loop, a unified event-triggered framework that jointly schedules analog and digital updates across orbital, channel, and traffic timescales remains an open research problem.
 

The prediction target defines the interface between AI inference and constrained HBF optimization. Beam-index prediction is compact but conveys limited information for multi-user interference suppression; full-CSI prediction incurs high acquisition overhead and sensitivity to channel aging; and direct beamformer generation reduces online optimization latency but requires explicit enforcement of constant-modulus, finite-resolution, RF-connectivity, and transmit-power constraints \cite{palacios_hybrid_2021,payami2018hybrid}. Lower-dimensional targets, including angular states, top-$K$ beam candidates, and effective CSI, offer a more practical compromise and are compatible with Kalman filtering, motion-aided prediction, learning-to-optimize, and deep-unfolding methods \cite{sim_unscented_2023,chiang_machine-learning_2021,huang2021massive,gao2022machine,sun2017learning,pellaco2021matrix,chowdhury2023deep}. However, existing ray-tracing, vision-, position-, and radar-aided datasets do not form a common NTN benchmark \cite{alkhateeb2019deepmimo,alrabeiah2020viwi,charan2022towards,liu2020location,demirhan2022radar}. Consequently, the end-to-end trade-off among prediction accuracy, signaling overhead, constrained beam synthesis, and update latency remains unresolved under identical mobility, array, codebook, pilot, and feedback-delay conditions.
 
A further challenge is bridging the gap between algorithm-level beam tracking and physically realizable aperture control. Most existing tracking schemes assume ideal array responses and continuous-resolution, constant-modulus phase control \cite{liu2020location, sim_unscented_2023, chiang_machine-learning_2021, gao2022machine}, whereas practical NTN payloads are affected by finite phase resolution, amplitude and phase mismatches, insertion loss, mutual coupling, and calibration drift, all of which can distort the realized beam independently of the state-prediction accuracy \cite{palacios_hybrid_2021, payami2018hybrid, chi_hybrid_2023}. Consequently, hardware-aware beam tracking must jointly model state-prediction error, constrained beam-synthesis error, and RF-realization error. Deployment realism also requires accounting for the platform on which inference is executed: reported UAV processing times, such as the approximately 11-ms latency in \cite{gao2022machine}, are measured on desktop computing hardware and may not represent the performance achievable by radiation-tolerant, power-constrained satellite or HAPS processors. Although several enabling components are available, including an NTN channel model for ns-3 \cite{sandri2023implementation}, the Sionna RT differentiable ray-tracing platform \cite{hoydis2023sionna}, a field-programmable gate array (FPGA)-based digital-beamforming receiver \cite{ortega2026digital}, GPU-accelerated AI-RAN tooling \cite{cammerer2025sionna}, and high-precision LEO beam-steering hardware \cite{chi_hybrid_2023}, they remain largely isolated. 
 
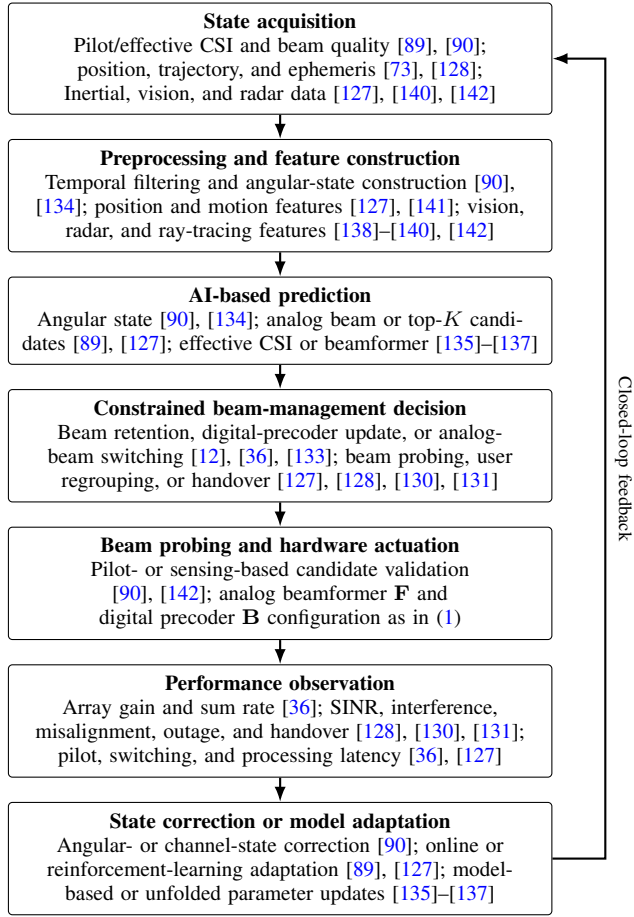
\begin{figure}[t]
\vspace{0.3cm}
\centering
\begin{tikzpicture}[
    node distance=3.2mm,
    block/.style={
        draw,
        rounded corners=1.5pt,
        align=center,
        text width=0.78\columnwidth,
        minimum height=8mm,
        inner sep=4pt,
        font=\footnotesize
    },
    flow/.style={
        -{Latex[length=2mm]},
        thick
    }
]
 
\node[block] (acquisition) {
    \textbf{State acquisition}\\
    Pilot/effective CSI and beam quality
    \cite{sim_unscented_2023,chiang_machine-learning_2021};
    position, trajectory, and ephemeris
    \cite{lin20215g,you_integrated_2024};
    Inertial, vision, and radar data
    \cite{gao2022machine,charan2022towards,demirhan2022radar}
};
 
\node[block, below=of acquisition] (preprocess) {
    \textbf{Preprocessing and feature construction}\\
    Temporal filtering and angular-state construction
    \cite{sim_unscented_2023,huang2021massive};
    position and motion features
    \cite{liu2020location,gao2022machine};
    vision, radar, and ray-tracing features
    \cite{alrabeiah2020viwi,charan2022towards,
    demirhan2022radar,alkhateeb2019deepmimo}
};
 
\node[block, below=of preprocess] (prediction) {
    \textbf{AI-based prediction}\\
    Angular state
    \cite{sim_unscented_2023,huang2021massive};
    analog beam or top-$K$ candidates
    \cite{chiang_machine-learning_2021,gao2022machine};
    effective CSI or beamformer
    \cite{sun2017learning,pellaco2021matrix,chowdhury2023deep}
};
 
\node[block, below=of prediction] (decision) {
    \textbf{Constrained beam-management decision}\\
    Beam retention, digital-precoder update, or analog-beam switching
    \cite{palacios_hybrid_2021,payami2018hybrid,momani_beam_2025};
    beam probing, user regrouping, or handover
    \cite{gao2022machine,lin20215g,zhu2024beam,zhu2024beam_2}
};
 
\node[block, below=of decision] (actuation) {
    \textbf{Beam probing and hardware actuation}\\
    Pilot- or sensing-based candidate validation
    \cite{sim_unscented_2023,demirhan2022radar};
    analog beamformer $\mathbf{F}$ and digital precoder
    $\mathbf{B}$ configuration as in \eqref{eq:tx}
};
 
\node[block, below=of actuation] (observation) {
    \textbf{Performance observation}\\
    Array gain and sum rate
    \cite{momani_beam_2025};
    SINR, interference, misalignment, outage, and handover
    \cite{lin20215g,zhu2024beam,zhu2024beam_2};
    pilot, switching, and processing latency
    \cite{gao2022machine,momani_beam_2025}
};
 
\node[block, below=of observation] (adaptation) {
    \textbf{State correction or model adaptation}\\
    Angular- or channel-state correction
    \cite{sim_unscented_2023};
    online or reinforcement-learning adaptation
    \cite{chiang_machine-learning_2021,gao2022machine};
    model-based or unfolded parameter updates
    \cite{sun2017learning,pellaco2021matrix,chowdhury2023deep}
};
 
\draw[flow] (acquisition) -- (preprocess);
\draw[flow] (preprocess) -- (prediction);
\draw[flow] (prediction) -- (decision);
\draw[flow] (decision) -- (actuation);
\draw[flow] (actuation) -- (observation);
\draw[flow] (observation) -- (adaptation);
 
\coordinate (feedbackright) at ($(adaptation.east)+(7mm,0)$);
\draw[flow]
    (adaptation.east)
    -- (feedbackright)
    |- node[pos=0.25, right, rotate=-90, anchor=south,
            font=\scriptsize]
       {Closed-loop feedback}
    (acquisition.east);
 
\end{tikzpicture}
\caption{Closed-loop workflow for AI-based beam tracking and management in HBF NTNs.}
\label{fig:ai_beam_tracking_workflow}
\vspace{-0.5cm}
\end{figure}
 
The overarching challenge is the end-to-end co-design of a closed-loop beam-management system that connects heterogeneous state acquisition, AI prediction, multi-timescale HBF decisions, hardware actuation, and feedback-based adaptation. As summarized in Fig. \ref{fig:ai_beam_tracking_workflow}, progress in any individual block is insufficient unless prediction uncertainty, RF impairments, onboard latency, and signaling overhead are jointly reflected in operational performance. A unified and experimentally validated workflow therefore remains central to translating AI-based beam tracking from isolated algorithms into practical NTN operation.
 
 
\subsubsection{Calibration and Mutual Coupling}

Calibration and mutual coupling are central to hardware-aware HBF in NTNs because practical apertures rarely realize the ideal analog response assumed in beamformer design. This issue becomes increasingly important for low-SWaP transmitarrays, lens-based beamformers, and reconfigurable apertures deployed on satellite and aerial platforms \cite{Song2026TransmitarrayReview}. For example, a recent gradient-index (GRIN)-lens HBF prototype required end-to-end calibration of the baseband--intermediate-frequency (IF)--mixer--switch chain together with explicit characterization of inter-port coupling \cite{Wen2026GRINLensHBF}. Similarly, passive TTD structures using Rotman lenses in \cite{Liu2018RotmanLens} mitigate beam squint through unequal propagation paths, but fabrication tolerances, port coupling, switch imbalance, and frequency-dependent loss perturb the intended delay profile, causing residual squint and leakage across beam ports.

These impairments add to amplitude and phase mismatches, quantized control states, non-uniform feed illumination, thermal drift, and aging, which distort the main beam, sidelobes, and spatial nulls \cite{Schmid2013CalibrationCoupling}. The RF-network topology also affects the realized response: conventional splitter--phase-shifter--attenuator--combiner networks incur intrinsic combining loss, whereas QR-factorization-based networks using tunable directional couplers can provide joint amplitude--phase control with lower loss \cite{sun2024novel}. Hence, calibration, coupling, insertion loss, and network topology need to be included directly in the HBF model rather than treated as post-design implementation effects.

Scanning reflectarray prototypes further expose the resulting observability problem. The hosted nimble beamforming anti-jam reflectarray (HoNi~BAJR) achieves low-SWaP adaptive beam shaping through free-space combining and reportedly reduces power consumption by approximately 95\% relative to a conventional active array \cite{gaines2026honi_bajr}. However, the absence of dedicated RF-chain access at each element limits direct observability and makes sidelobe and null control highly sensitive to small hardware variations. The remaining challenge is therefore an online calibration and beam-synthesis framework that jointly estimates the realized aperture response and updates beam steering, null shaping, and interference suppression under limited measurements and time-varying hardware drift.
 
\subsection{Integrated Sensing, Communication, and Beamforming (ISAC-HBF)}
As illustrated in Fig. \ref{fig:NTN_ISAC_signal}, sensing and HBF form a closed loop in which echoes, angles, delays, and Doppler support beam tracking, CSI acquisition, and localization, while the resulting analog and digital beams jointly serve users and illuminate targets. Existing studies have investigated this interaction through wideband LEO HBF, communication–localization tradeoffs, RSMA, sensing-assisted UAV control, and RIS-assisted links \cite{you_beam_2022, you_integrated_2024, liu_hybrid_2025, zhang_sensing-assisted_2024, wu_joint_2025, salih_spectral_2025}. Recent work has further extended ISAC-HBF from single-platform monostatic systems toward robust multibeam transmission, bistatic sensing, predictive handover, and distributed constellation cooperation \cite{cao_robust_2025, park2025bistatic, bhandari2026isac, zhang2026distributed, zhang2026leo}, thereby motivating three major challenges.
 
\subsubsection{Robust and Mobility-Aware Hybrid Beamforming}
Robust ISAC-HBF must translate correlated errors in angle, delay, Doppler, attitude, calibration, and feedback into joint communication-and-sensing beamformer constraints. Existing methods exploit statistical CSI, geometry, ephemeris, or estimated angles \cite{you_beam_2022, you_integrated_2024, liu_hybrid_2025, zhang_sensing-assisted_2024, park2025bistatic}, but generally  assume independent errors. The ISAC-specific open problem is multi-rate state estimation and uncertainty-aware analog-beam tracking that preserves communication, detection, tracking, and localization performance while fast digital refinement operates between analog updates \cite{zhang_sensing-assisted_2024, salih_spectral_2025}.
 
\subsubsection{Wideband Hardware-Constrained Hybrid Beamforming}
Wideband ISAC-HBF faces frequency-dependent mismatch between communication and sensing beams. Although large bandwidths improve data rate and delay resolution, frequency-flat phase shifters cause beam squint across subcarriers \cite{you_beam_2022, you2024ubiquitous}. TTD-assisted architectures mitigate this effect, but finite delay ranges, quantized phase control, limited RF chains, and hardware power consumption constrain the achievable beampatterns. Moreover, communication favors narrow user-directed beams, whereas sensing may require broader or scanning beams. These conflicting requirements motivate joint design of waveforms, subcarrier allocation, analog connectivity, TTD configurations, and digital precoders under practical hardware constraints \cite{you_beam_2022, liu_hybrid_2025}.
 
\subsubsection{Real-Time Implementation and End-to-End Validation}
Existing ISAC-HBF designs rely on multi-stage iterative solvers, including alternating optimization for wideband LEO HBF \cite{you_beam_2022}; PDD combined with WMMSE, sequential rank-one constraint relaxation (SROCR), and the concave-convex procedure (CCP) for RSMA-enabled LEO-ISAC \cite{liu_hybrid_2025}; AO integrating SDR, manifold optimization, and SCA for sensing-assisted UAV networks \cite{zhang_sensing-assisted_2024}; WMMSE--PDD--AO with SDR or majorization-minimization (MM) for active-RIS-assisted DFRC \cite{wu_joint_2025}; and two-layer SDR--SROCR--SCA optimization for bistatic LEO-ISAC \cite{park2025bistatic}. Riemannian manifold optimization and multi-agent DRL have also been applied to ISAC-assisted handover \cite{bhandari2026isac}, while PDD-based two-layer optimization has been adopted for bistatic HBF \cite{zhang2026leo}. These methods are evaluated primarily offline, leaving their ability to satisfy onboard sensing and beam-update deadlines unclear. Further research requires latency- and resource-aware implementations together with end-to-end evaluation over complete satellite passes or UAV trajectories, covering communication reliability, sensing accuracy, localization error, processing latency, beam-training overhead, and energy consumption.
 
\subsection{3D Multi-Layer NTN}\label{ssec:3dntn}

\begin{figure*}[!t]
\centering
\includegraphics[width=\linewidth]{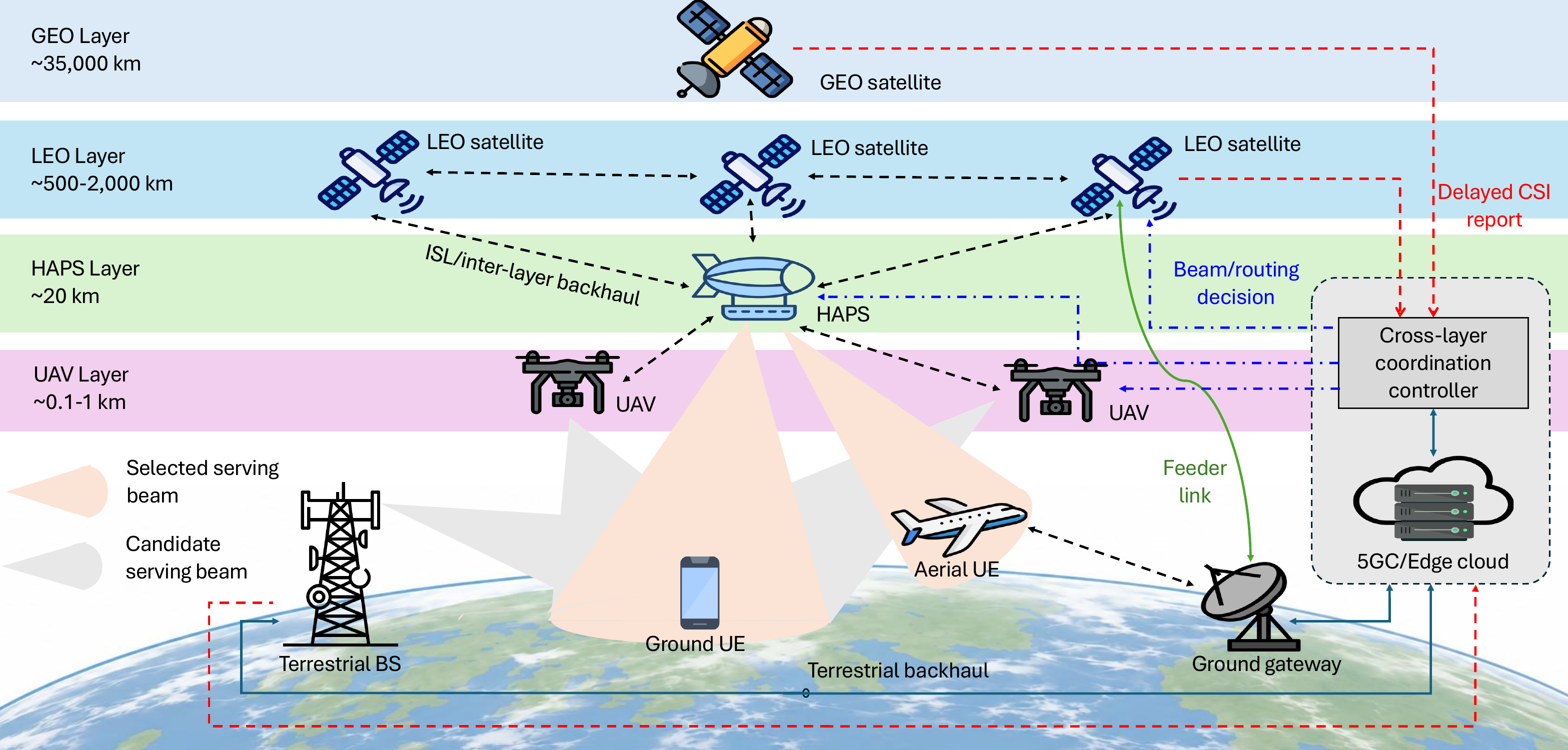}
\caption{Conceptual architecture of a 3D multi-layered NTN for cross-layer HBF design.}
\label{fig:3d_multi_layered NTN}
\vspace{-0.25cm}
\end{figure*}
 
As illustrated in Fig. \ref{fig:3d_multi_layered NTN}, the considered 3D multi-layered NTN integrates GEO and LEO satellite layers, a HAPS layer, a UAV layer, terrestrial infrastructure, a ground gateway, and ground and aerial UEs. The selected and candidate beam footprints represent alternative service links from terrestrial and aerial platforms, whereas the LEO layer primarily supports inter-satellite links (ISLs), LEO–HAPS backhaul, and feeder connectivity to the gateway. Direct LEO–UE access can be incorporated as an additional option but is not explicitly shown in the illustrated baseline. A cross-layer coordination controller collects delayed CSI and network-state reports and returns beam and routing decisions to the controlled platforms. Unlike single-layer HBF, this architecture coordinates nodes with different propagation delays, mobility patterns, coverage footprints, antenna apertures, RF-chain resources, and onboard processing capabilities \cite{geraci2022integrating, scalised3, rago2024multi, khennoufa2025multi}. The overlapping service regions and coupled access–backhaul topology consequently make beamforming dependent on user association, multi-connectivity, handover, and the availability of inter-layer transport resources.
 
For a UE covered by multiple selected or candidate beams, the design problem involves jointly determining the serving platform, active service links, analog and digital beamformers, power and traffic allocation, and handover timing. These variables are coupled through platform-dependent propagation and mobility, array and RF-chain resources, backhaul conditions, and nonideal RF hardware, which may render a strong access link infeasible or inefficient \cite{khennoufa2025multi}. Existing multi-layer scheduling and slicing studies demonstrate the benefits of multi-connectivity but typically rely on simplified visibility, terminal-capability, or network-state assumptions and do not jointly address rapid beam evolution and handover \cite{dazhi2023energy, dazhi2024joint}. Association is therefore more appropriately formulated as an end-to-end, HBF-aware problem rather than one based only on received power or instantaneous SINR. For example, a high-SINR HAPS beam may be suboptimal when its LEO backhaul is congested, no RF chain is available, or an imminent handover would cause excessive interruption. Although candidate-beam activation can support make-before-break operation, it also increases signaling, RF-chain usage, power consumption, and interference. These couplings motivate a unified framework for joint association, HBF, traffic splitting, and beam/platform handover under hardware, backhaul, and beam-continuity constraints.
 
While joint association, HBF, and handover determine the preferred platforms, links, and beamformers, their practical implementation depends on timely and sufficiently accurate network-state information. As shown in Fig. \ref{fig:3d_multi_layered NTN}, the cross-layer controller receives delayed CSI and network reports and returns beam and routing decisions to heterogeneous platforms. However, reporting delays differ across terrestrial, UAV, HAPS, LEO, and GEO segments, while ISLs and inter-layer links have limited signaling capacity. By the time a centralized HBF solution is delivered, the satellite geometry, UAV attitude, user position, or interference conditions may have changed. Heterogeneous Doppler, propagation delays, array responses, and hardware constraints further limit centralized perfect-CSI designs \cite{geraci2022integrating, scalised3, rago2024multi, khennoufa2025multi}.
 
These limitations motivate hierarchical coordination, where the cross-layer controller handles slower association, traffic-splitting, backhaul-aware allocation, and handover decisions, while local nodes perform fast beam tracking, digital-precoder refinement, and beam-failure recovery. Exchanging beam indices, predicted angular states, effective CSI, and resource indicators instead of full CSI and HBF matrices can reduce signaling overhead. Nevertheless, the required information, its update rate, and the enforcement of end-to-end constraints from inconsistent local observations remain unresolved. Predictive HBF may compensate for control delay, but the appropriate prediction target, such as beam index, angular state, effective CSI, or hardware-feasible beamformer, and its accuracy–overhead trade-off across NTN layers require further investigation. Evaluation over complete satellite passes and dynamic aerial trajectories is also needed to capture CSI aging, coordination overhead, inference latency, and beam-failure recovery.
 
\subsection{THz HBF for NTN}
Current research on THz-enabled NTNs spans wideband hybrid precoding, energy-efficient antenna architectures, propagation characterization, and RIS-assisted transmission. The angle-based ITS architecture in \cite{wu_hybrid_2023} combines frequency-independent analog control with subcarrier-dependent digital precoding to mitigate beam squint, while \cite{mir_hybrid_2020} develops a closed-form sum-MSE hybrid precoder for sub-connected THz relay systems. Complementary studies characterize THz transceiver and circuit constraints~\cite{10918779} as well as transmitarray antenna designs~\cite{Song2026TransmitarrayReview}, atmospheric and mobility effects across satellite, HAPS, UAV, and ground links \cite{amodu2024technical, erdem2026terahertz}, and the combined impact of turbulence, pointing errors, and finite-resolution phase control \cite{yuan2022secure}. Building on these developments, future research can focus on atmospheric-window-aware precoding, mixed near-/far-field wideband focusing, and spatial-rank-aware design of ultra-massive HBF architectures.
 
\subsubsection{Frequency-Window-Aware Spatial–Spectral HBF}
Molecular resonances divide the THz band into transmission windows separated by absorption peaks, with loss determined by frequency, altitude, humidity, distance, and link geometry \cite{10918779, amodu2024technical, erdem2026terahertz}. High-altitude HAPS-to-HAPS and HAPS-to-satellite paths experience a smaller atmospheric contribution than HAPS-to-ground and HAPS-to-UAV paths \cite{erdem2026terahertz}. This response couples spectrum selection to HBF: the analog aperture is shared across the assigned band, while absorption, molecular noise, and received SNR vary across subcarriers. The per-subcarrier digital compensation in \cite{wu_hybrid_2023} reduces spatial mismatch but does not select the occupied atmospheric windows. The open problem is a joint design of transmission windows, subcarriers, bandwidth, power, analog coefficients, and digital precoders under subcarrier-resolved absorption and noise models. For multiuser links, different atmospheric paths create user-dependent feasible subbands, which further couples beam formation to beam–frequency assignment.
 
\subsubsection{Wideband Near-/Far-Field Beam Focusing}
Satellite and long-range HAPS links occupy the far field under many aperture choices, whereas nearby aerial terminals and large RIS/ITS-assisted segments enter the radiative near field. The resulting NTN contains planar-wave, spherical-wave, and transition-region links \cite{10918779, amodu2024technical, erdem2026terahertz}. Far-field HBF controls angle, while near-field HBF controls angle and range; wideband operation adds frequency-dependent direction and focal-distance shifts. The angle-based ITS design in \cite{wu_hybrid_2023} addresses far-field squint but not spherical wavefronts or focal displacement. Research is needed on polar-domain channel models, angle–range–frequency codebooks, mixed-field user multiplexing, and hybrid precoders that preserve focal gain without enumerating a dense three-dimensional training grid.
 
\subsubsection{Ultra-Massive Arrays under Limited Spatial Degrees of Freedom}
THz wavelengths accommodate hundreds or thousands of elements in a compact aperture, providing coherent gain against propagation loss; however, LoS dominance and weak reflected paths keep the channel spatial rank small \cite{10918779, mir_hybrid_2020}. Element count, therefore, increases array gain without a proportional increase in multiplexing capability. Fully connected HBF scales RF connectivity with antennas and RF chains, while fixed sub-connected processing reduces connectivity at the cost of analog-subspace flexibility. Existing work offers partial solutions. For example, the work \cite{wu_hybrid_2023} uses four active feeds to illuminate a $16 \times 16$ ITS, and the authors in \cite{mir_hybrid_2020} apply block-diagonal sub-connected processing at a two-way relay. Neither selects the architecture from channel rank, propagation regime, or NTN geometry. The open challenge is joint optimization of array geometry, subarray partitioning, feed placement, RF-chain count, and analog–digital dimensions, using the number of usable spatial modes per unit of aperture and hardware power as the principal design criterion.
 

\section{Concluding Remarks}\label{Sect:Conclusion}
This survey has treated HBF as the common physical-layer foundation of two non-terrestrial platform families usually studied in isolation: LEO satellite and UAV communication systems. From shared fundamentals on signal models, sparse channels, analog and digital precoding, and learning-aided methods, we organized the literature along a platform-, architecture-, and methodology-aware taxonomy, that applies the same five categories to both platform families, namely system architecture and precoding design, time-varying beam management, network-level design, sensing capability and reconfigurable surfaces, and security and multiple access, positioning ISAC, RIS and RHS, THz operation, security, and NOMA and RSMA access as cross-cutting extensions rather than isolated subjects.

The recurring lesson is that HBF architecture and algorithm must be co-designed with the platform, not selected after it. An LEO payload couples to quantities it can predict but not alter, i.e., orbital trajectory, Doppler, beam squint, and channel aging, favoring predicted CSI and a split between slow analog tracking and fast digital adaptation. A UAV payload couples to quantities it controls, i.e., altitude, position, and trajectory, making placement a first-class beamforming variable and favoring joint trajectory--beamforming co-design. Common to both, CSI quality matters more than quantity, robustness is platform-specific, sensing serves as a practical CSI surrogate, and learning is most credible when anchored to physical structure.


Despite substantial progress, HBF for NTNs remains validated primarily through simulation, studied largely in single-platform settings, and modeled with idealized RF hardware. Addressing the challenges identified in Section~\ref{Sect:OpenChallenges}, therefore, requires end-to-end, platform-aware co-design in which wideband operation, RF impairments, prediction uncertainty, processing latency, and signaling overhead are evaluated jointly. Experimental validation over realistic satellite passes and UAV trajectories will be essential for translating isolated algorithmic advances into deployable NTN systems.

\balance
\bibliographystyle{IEEEtran}
\bibliography{reference}

\end{document}